\documentclass[a4paper,11pt]{article}
\usepackage{jheppub} % for details on the use of the package, please see the JINST-author-manual
\usepackage[utf8]{inputenc}
\usepackage{graphicx}
\usepackage{amsmath}
\allowdisplaybreaks[1] 
\usepackage{hyperref}
\usepackage{cancel}
\usepackage{multirow}
\usepackage{slashed}
\usepackage{amssymb}
\usepackage{booktabs}

\arxivnumber{2411.09578} % if you have one

\title{\boldmath Constraints on the hadronic light-by-light tensor in corner kinematics for the muon $g-2$}

\author[a]{J.~Bijnens}
\author[a,b]{N.~Hermansson-Truedsson}
\author[c]{A.~Rodr\'{i}guez-S\'{a}nchez}
\affiliation[a]{Division of Particle and Nuclear Physics, Department of Physics, Lund University,
Box 118, \\SE 221 00 Lund, Sweden}
\affiliation[b]{Higgs Centre for Theoretical Physics, School of Physics and Astronomy, The University of Edinburgh, James Clerk Maxwell Building,
Peter Guthrie Tait Road,
Edinburgh,
EH9 3FD}
\affiliation[c]{Departament de F\'{i}sica Te\'{o}rica, IFIC, Universitat de Val\`{e}ncia CSIC, 
Apt. Correus 22085, \\E-46071 Val\`{e}ncia, Spain}

\emailAdd{johan.bijnens@fysik.lu.se}
\emailAdd{nils.hermansson-truedsson@ed.ac.uk}
\emailAdd{anrosanz@ific.uv.es}

\abstract{The dispersive approach to the hadronic light-by-light contribution to the muon $g-2$ involves an integral over three virtual photon momenta appearing in the light-by-light tensor. Building upon previous works, we systematically derive short-distance constraints in the region where two momenta are large compared to the third, the so-called Melnikov-Vainshtein or corner region. We include gluonic corrections for the different scalar functions appearing in the Lorentz decomposition of the underlying tensor, and explicitly check analytic agreement with alternative operator product expansions in overlapping regimes of validity. A very strong pattern of cancellations is observed for the final $g-2$ integrand. The last observation suggests that a very compact expression only containing the axial current form factors can provide a good approximation of the corner region of the hadronic light-by-light tensor.}

\begin{document}
\maketitle
\flushbottom

\section{Introduction}
\label{sec:introduction}
\begin{figure}
    \begin{center}
     \includegraphics[width=0.25\textwidth]{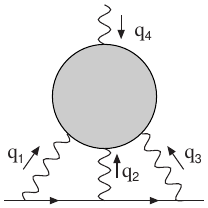}   
    \end{center}
    \caption{\label{fig:hlbl}The HLbL contribution to the $(g-2)_\mu$.}
\end{figure}

The experimental programme measuring the muon anomalous magnetic moment, or muon $g-2$, at Fermilab is close to completion~\cite{Muong-2:2015xgu,Muong-2:2021ojo,Muong-2:2023cdq}. In order to take full advantage of the experimental progress, the theoretical uncertainty on the Standard Model prediction of the $g-2$ has to be reduced to match the size of the final experimental uncertainty. The most recent theoretical Standard Model prediction for $a_\mu = (g-2)_\mu/2$ can be found in the White Paper~\cite{Aoyama:2020ynm}, and builds on data-driven approaches as well as lattice quantum chromodynamics (QCD). 
For the hadronic light-by-light (HLbL) contribution, diagrammatically depicted in Fig.~\ref{fig:hlbl}, the data-driven approach leads to \cite{Aoyama:2020ynm,Melnikov:2003xd,Masjuan:2017tvw,Colangelo:2017fiz,Hoferichter:2018kwz,Gerardin:2019vio,Bijnens:2019ghy,Colangelo:2019uex,Pauk:2014rta,Danilkin:2016hnh,Jegerlehner:2017gek,Knecht:2018sci,Eichmann:2019bqf,Roig:2019reh}
\begin{align}
    a_{\mu}^{\textrm{HLbL}} = 92(19)\times 10^{-11} \, ,
\end{align}
which has a relative uncertainty at the 20\% level. The precision goal for the HLbL contribution is an uncertainty of 10\%. Several recent works in the data-driven approach have appeared with the goal of reducing the uncertainty, see e.g.~Refs.~\cite{Colangelo:2021nkr,Bijnens:2020xnl,Bijnens:2021jqo,Bijnens:2022itw,Leutgeb:2019gbz,Cappiello:2019hwh,Masjuan:2020jsf,Ludtke:2020moa,Leutgeb:2021mpu,Leutgeb:2021bpo,Hoferichter:2020lap,Knecht:2020xyr,Zanke:2021wiq,Danilkin:2021icn,Cappiello:2021vzi,Hoferichter:2023tgp,Ludtke:2023hvz,Colangelo:2024xfh,Hoferichter:2024fsj,Deineka:2024mzt,Ludtke:2024ase,Estrada:2024cfy,Holz:2024lom,Eichmann:2024glq,Leutgeb:2024rfs,Holz:2024diw,Hoferichter:2024vbu,Hoferichter:2024bae,Cappiello:2025fyf}, and must be combined with the results here for a future improved prediction of $a_{\mu}^{\textrm{HLbL}}$~\cite{Colangelo:2022jxc}. Lattice QCD offers an alternative high-precision determination of the HLbL contribution, see e.g.~Refs.~\cite{Blum:2019ugy,Chao:2021tvp,Chao:2022xzg,Blum:2023vlm,Fodor:2024jyn}, with further progress also expected in the near future. 

A significant part of the current uncertainty on the data-driven HLbL (roughly $ 10\times 10^{-11}$) comes from short-distance contributions. An integral part of controlling the uncertainty from the short-distance contributions is the use of constraints that can be derived systematically using the operator product expansion (OPE)~\cite{Melnikov:2003xd,Bijnens:2019ghy,Bijnens:2020xnl,Bijnens:2021jqo,Bijnens:2022itw}. For the HLbL different expansions are needed for the various kinematical configurations of the three virtual photons in Fig.~\ref{fig:hlbl}. 

In this paper, we continue our recent work on short-distance expansions in the limit where two of the momenta in the integration\footnote{The momenta are the off-shellness of the three internal photons in Fig.~\ref{fig:hlbl}.} are large compared to the third.
In Ref.~\cite{Bijnens:2022itw} we generalized the Melnikov-Vainshtein corner OPE result of Ref.~\cite{Melnikov:2003xd} to next-to-leading order (NLO) in the power expansion and showed how the systematic power counting works for the different scalar functions involved in the $g-2$ integral of Refs.~\cite{Colangelo:2015ama,Colangelo:2017fiz}. We found that most of the scalar functions, $\bar{\Pi}_i$, actually starts at NLO in the power expansion (driven by dimension $D=4$ operators), and recovered the quark loop result for the kinematic regime where it is valid.

In this work we go one step further. First we also compute gluonic corrections, showing that when one can compute the matrix elements perturbatively, one fully recovers order by order in the OPE the two-loop results of Ref.~\cite{Bijnens:2021jqo}, valid for three large virtual momenta. This is done in Sec.~\ref{sec:OPEresults}, after a presentation of the overall formalism and notation in Sec.~\ref{sec:notation}. 
Second, in Sec.~\ref{sec:amucancellations} we combine the obtained expanded scalar functions with the $g-2$ integration kernels and find a very strong pattern of cancellations when adding the different scalar functions together, leading to compact expressions for the corresponding $g-2$ contributions. Conclusions and an outlook are presented in Sec.~\ref{sec:conclusions}. A number of technical aspects are relegated to the appendices.

\section{The hadronic light-by-light tensor and $a_\mu^{\textrm{HLbL}}$}
\label{sec:notation}

The central object needed is the HLbL tensor, defined as the correlation function of four electromagnetic currents $J^{\mu}(x) = \sum _q e_q\, \bar{q}(x)\gamma^{\mu}q(x)$ with $q=u,d,s$, $e_u = 2/3$ and $e_d = e_s = -1/3$. Following the notation of Ref.~\cite{Bijnens:2019ghy}, the HLbL tensor is then
\begin{equation}\label{eq:hlbltensor}
\Pi^{\mu_{1}\mu_{2}\mu_{3}\mu_{4} } 
=
-i\int \frac{d^{4}q_{4}}{(2\pi)^{4}}\left(\prod_{i=1}^{4}\int d^{4}x_{i}\, e^{-i q_{i} x_{i}}\right)  \langle 0 | T\left(\prod_{j=1}^{4}J^{\mu_{j}}(x_{j})\right)|0\rangle \, .
\end{equation}
The four momenta satisfy $q_1+q_2+q_3+q_4 = 0$, and the HLbL tensor obeys the set of four Ward identities
\begin{align}
    q_{i\, \mu _i}\Pi^{\mu_{1}\mu_{2}\mu_{3}\mu_{4} } = 0 \, . 
\end{align}
From these Ward identities it follows that we may write the tensor entirely in terms of its derivative in $q_4$,
\begin{align}\label{eq:wardcons}
\Pi^{\mu_{1}\mu_{2}\mu_{3}\mu_{4}}=-q_{4\, \nu_{4}}\frac{\partial \Pi^{\mu_{1}\mu_{2}\mu_{3}\nu_{4}}}{\partial q_{4\, \mu_{4}}}\, .
\end{align}
The HLbL contribution to the muon anomalous magnetic moment, $a_\mu^{\textrm{HLbL}}$, is depicted in Fig.~\ref{fig:hlbl}, where the external photon with momentum $q_4$ is soft, i.e.~where $q_4\rightarrow 0$. As shown in Refs.~\cite{Colangelo:2015ama,Colangelo:2017fiz} $a_\mu^{\textrm{HLbL}}$ is given by\footnote{The choice of variables and how to split the HLbL tensor into scalar functions is not unique, we will use the conventions of Refs.~\cite{Colangelo:2015ama,Colangelo:2017fiz}. Older versions are e.g.~in Ref.~\cite{Bijnens:1995xf}.} 
\begin{align}\label{eq:amuint}
a_{\mu}^{\textrm{HLbL}} 
= 
\frac{2\alpha ^{3}}{3\pi ^{2}} 
& \int _{0}^{\infty} dQ_{1}\int_{0}^{\infty} dQ_{2} \int _{-1}^{1}d\tau \, \sqrt{1-\tau ^{2}}\, Q_{1}^{3}Q_{2}^{3}
\sum _{n=1}^{12} T_{n}(Q_{1},Q_{2},\tau)\, \overline{\Pi}_{n}(Q_{1},Q_{2},\tau)\, .
\end{align}
Here the variables $Q_{i}$ are the magnitudes of the Euclidean squares of the momenta $q_i$, $Q_i^2 = -q_i^2$, and $\tau$ relates the three scales through $Q_3^2=Q_1^2+Q_2^2+2\tau Q_1Q_2$. The kernels $T_i $ are known functions given in Ref.~\cite{Colangelo:2015ama}. The $\overline{\Pi}_{i}$ can be determined from a set of 6 functions $\hat{\Pi}_j$ determined by the derivative of the HLbL tensor in Eq.~(\ref{eq:wardcons}). Knowing the derivative tensor, we may project onto the $\hat{\Pi}_i$ as described in Refs.~\cite{Bijnens:2020xnl,Bijnens:2021jqo,Bijnens:2022itw}.  

An alternative formulation of the above integral in terms of the variables $Q_1,Q_2,Q_3$ is
\begin{align}
a_{\mu}^{\textrm{HLbL}} 
& 
=  \frac{2\alpha ^{3}}{3\pi ^{2}} 
 \int_0^\infty dQ_1    \int_0^\infty dQ_2   \int_0^\infty dQ_3 \, \frac{1}{2}\, Q_1 Q_2 Q_3\, \sqrt{-\lambda(Q_1^2, Q_2^2, Q_3^2)} 
 \nonumber \\
 & \times \mathcal{R}(Q_1,Q_2,Q_3)\, \sum _{n=1}^{12} T_{n}(Q_{1},Q_{2},\tau)\, \overline{\Pi}_{n}(Q_{1},Q_{2},\tau) \, . 
\end{align}
Here $\mathcal{R}(Q_1,Q_2,Q_3)$ restricts $Q_1,Q_2,Q_3$ to form a triangle, i.e. can be expressed in terms of the Heaviside step function as $\mathcal{R}(Q_1,Q_2,Q_3)=\Theta(-\lambda (Q_1^2,Q_2^2,Q_3^2))$, where the K\"{a}ll\'{e}n function is defined as $\lambda (Q_1^2,Q_2^2,Q_3^2) = (Q_1^2+Q_2^2-Q_3^2)^2 - 4\, Q_1^2Q_2^2 $. The functions $T_n$ and $\overline{\Pi}_n$ should be understood as the ones in Eq.~(\ref{eq:amuint}) but with $\tau$ expressed in terms of $Q_1$, $Q_2$ and $Q_3$.

For our purposes it is convenient to define new variables referred to as corner variables. We introduce $\overline{Q}_i=Q_j+Q_k$ and $y_{jk}=\frac{Q_j-Q_k}{Q_i}$ which gives the integral
\begin{align}
a_\mu^{\textrm{HLbL}}&=a_\mu^{\textrm{HLbL},i}=\frac{2\alpha ^{3}}{3\pi ^{2}}\int_0^\infty d\overline Q_i    \int_0^{\overline Q_i}dQ_i   \int_{-1}^1 dy_{jk} \,
   \frac{1}{16}\left({\overline Q_i}^2-y_{jk}^2Q_i^2\right) Q_i^3\sqrt{{\overline Q_i}^2- Q_i^2} 
   \nonumber \\
   & \qquad \quad 
   \times 
   \sqrt{1-y_{jk}^2}
   \, \sum_{n=1}^{12} T_n \left(\overline{Q}_i,Q_i,y_{jk}\right) \, \overline{\Pi}_n \left(\overline{Q}_i,Q_i,y_{jk}\right) \, .
\end{align}
These exist for $ijk=312,123,231$.
The functions $T_n$ and $\overline{\Pi}_n$ should again be understood as the ones in Eq.~(\ref{eq:amuint}), but now with $Q_{1}$, $Q_2$ and $\tau$ expressed in terms of the relevant corner variables. 

Separating long and short distances can be done by introducing a scale $\Lambda$ separating a long and short distance for each of $Q_1,Q_2,Q_3$. A discussion of the regions in the different integration variables is given in appendix~\ref{app:kinematics}. In particular it should be noted that all integration measures are always even in the $y_{jk}$ variables for each of the regions and the regions can be combined such that the integration is also even in $y_{jk}$.

\section{The two-current OPE}
\label{sec:OPEresults}

We next proceed to consider an OPE of the HLbL tensor $\Pi^{\mu_{1}\mu_{2}\mu_3\mu_4}$. As we will be interested in the mixed Melnikov-Vainshtein kinematics we must then decide an ordering among the photon virtualities. We here consider the case where $Q_3$ is small compared to $Q_1$ and $Q_2$, but stress that the other cases can be treated analogously. We start by rewriting the HLbL tensor with two external fields associated with the photons with momenta $q_3$ and $q_4$ according to~\cite{Bijnens:2022itw}
\begin{align}\label{eq:twopoint}
\Pi^{\mu_{1}\mu_{2}\mu_3\mu_4}=
&
\sum_{j,k}\frac{i e_{q_j}e_{q_k}}{e^{2}}\int \frac{d^{4}q_4}{(2\pi)^4}\int d^{4}x_1\int d^{4}x_2\, e^{-i(q_1 x_1+q_2 x_2)}
\nonumber \\
& \times 
\langle 0 |T\{ J_j^{\mu_1}(x_1)J_k^{\mu_2}(x_2)\}|\gamma^{\mu_3}(q_3)\gamma^{\mu_4}(q_4) \rangle \, .
\end{align}
This should be understood through the Dyson series,\footnote{The Dyson series give extra factors of $e$ from the extra $e\, A_{\mu}(x)J^{\mu}(x)$ interaction terms. These cancel against the $1/e^2$ pre-factor, which is implemented to recover the same order in $\alpha$ as in Eq.~(\ref{eq:hlbltensor}).} where vertices containing the photon field $A^{\alpha}(x)$ contracted with an external state (with Lorentz index $\mu$) yield
\begin{equation}
A^{\alpha}(x)|\gamma^{\mu}(q)\rangle\equiv g^{\alpha\mu}e^{-iqx} \, .
\end{equation}
We will repeatedly refer to the flavor octet and singlet pieces, defined through $ \Pi^{\mu_1\mu_2\mu_3\mu_4} = \Pi^{\mu_1\mu_2\mu_3\mu_4}_{(8)}+\Pi^{\mu_1\mu_2\mu_3\mu_4}_{(1)}$, where\footnote{The notation $\langle A \rangle ^{j,\mu _3 ,\mu _4}_{q_4,x_1,x_2}  = \int d^4 q_4 \, d^4 x_1 \, d^4 x_2 \, \langle 0 | A_{j}| \gamma ^{\mu _3}(q_3) \gamma ^{\mu _4}(q_4)\rangle$, where again $j$ is a flavor index. }
\begin{align}
    \Pi^{\mu_1\mu_2\mu_3\mu_4}_{(F)} = 
    \sum_j \mathcal{T}^{(F)}_{j}
    \, 
\Big\langle e^{-i(q_1 x_1 +q_2 x_2)}\, \bar{q}(x_1)\gamma^{\mu_1}q(x_1) \;\bar{q}(x_2)\gamma^{\mu_2}q(x_2) \Big\rangle ^{j,\mu_3,\, \mu_4}_{q_{4},\, x _1,\, x_2} \, ,
\end{align}
with the flavor projectors
\begin{align}
   \mathcal{T}^{(8)}_{j} 
   & 
   = \frac{i  (e_{q_j}^{2}-\sum_{k} \frac{e_{q_k}^2}{3})}{e^2}
   \, ,
   \\
  \mathcal{T}^{(1)}_{j}   
    & 
    =
    \frac{i \sum_{k} \frac{e_{q_k}^2}{3}}{e^2}
    \, . 
\end{align}

We further define the variables
\begin{align}
    \hat q &= \frac{1}{2}\left(q_1-q_2\right)\, , & q_{1,2} &= \pm\hat q-\frac{1}{2}(q_3+q_4)\, ,
\end{align}
and the Euclidean large scale $\hat{Q} = \sqrt{-\hat{q}^2}$. As in our earlier work (Ref.~\cite{Bijnens:2022itw}) we may perform the OPE in terms of the large momentum $\hat{Q}$ (as compared to $q_3$, $q_4$ and $\Lambda_{\mathrm{QCD}}$). We will review the main results of this OPE below, and also extend it to include order-$\alpha _s$ corrections. It is important to note that the expansion makes no assumption on the inherent ordering between $Q_3$ and $\Lambda _{\textrm{QCD}}$, i.e.~$Q_3$ can be in the perturbative or non-perturbative regime.  

In order to recover the OPE result obtained in Ref.~\cite{Bijnens:2022itw}, one can start by connecting the two quark current insertions in Eq.~(\ref{eq:twopoint}) to then add as many perturbative QCD (and QED) vertices as needed up to the studied order. Once the external momenta flowing through the remaining legs are taken to be small, the corresponding contributions can be written as a sum of perturbative Wilson coefficients multiplying matrix elements of local operators.

At the studied order in $\alpha_s$, neglecting contributions from possible $D=4$ operators that vanish in the chiral limit, the bare operators through dimension $D=4$ that appear in the matrix elements are\footnote{We drop the quark-flavor index $j$ here on the quark fields. The two combinations of operators $\mathcal{O}_{F}$ and $ \mathcal{O}_{G} $ are sets of operators with other Lorentz structures than $\mathcal{O}_{FF,i}^{\alpha  \beta}$ and $\mathcal{O}_{GG,i}^{\alpha  \beta}$ (see bottom line of Eq.~(\ref{eq:masteroperen})). %Additional combinations of operators with two $F$ or two $G$ (but with finite Wilson coefficients) also enter in the calculation. See bottom line of Eq.~(\ref{eq:masteroperen}).
}
\begin{align}
			D=3: & \quad \mathcal{O}_{q3}^{\alpha \beta \gamma} =  \overline{q} \Big[ 
			\gamma ^{\alpha}\gamma ^{\beta} \gamma ^{\gamma}
			-
			\gamma ^{\gamma}\gamma ^{\beta} \gamma ^{\alpha} 
			\Big] q
     \, ,
	\\
			D=4: &  \quad \mathcal{O}_{q4}^{\alpha \beta} = \overline{q} 
			\gamma ^{\beta}
			\Big[ 
			\overrightarrow D^\alpha
   -
   \overleftarrow D^\alpha
			\Big] q 
     \, ,
			\\
			&  \quad \mathcal{O}_{FF,1}^{\alpha  \beta} = F^{\alpha \gamma}F_{\gamma }^{\, \,  \beta}
     \, ,
			\\
			&  \quad \mathcal{O}_{FF,2}^{\alpha  \beta} = F^{\gamma \delta }F_{\gamma \delta} \, g^{\alpha \beta}
     \, ,
			\\
			& \quad 
             \mathcal{O}_{F} = F\times F
              \, ,
			\\
			& \quad 
            \mathcal{O}_{GG,1}^{\alpha  \beta} = G^{\alpha \gamma}G^{\, \, \beta}_{\gamma}
     \, ,
			\\
			& \quad \mathcal{O}_{GG,2}^{\alpha  \beta} = G^{\gamma \delta }G_{\gamma \delta} \, g^{\alpha \beta}
   \, ,
   \\
			& \quad 
             \mathcal{O}_{G} = G\times G
              \, ,
			\\
   \label{eq:redundantop}
			& \quad \tilde{\mathcal{O}}_{q4}^{\alpha  \beta} =  \overline{q} \Big[
				\gamma ^{\alpha}\gamma ^{\gamma} \gamma ^{\beta}
			+
			\gamma ^{\beta}\gamma ^{\gamma} \gamma ^{\alpha} 
				\Big] \Big[ \overrightarrow{D}_{\gamma} + \overleftarrow{D}_{\gamma}
				\Big] q \, .
\end{align}
We note that the operators involve quark fields as well as the field-strength tensors $F_{\mu \nu}$ and $G_{\mu \nu}^{a}$. The last operator is redundant, since by using the quark equations of motion it can be reduced to $\mathcal{O}_{q4}^{\alpha\beta}$. It is also a total
derivative of $\mathcal{O}_{q3}^{\alpha\beta\gamma}$. The precise relation is
\begin{align}\label{eq:eom}
    \widetilde{\mathcal{O}}_{q4}^{\alpha\beta} = D_\gamma \mathcal{O}_{q3}^{\alpha\beta\gamma}
    = 2\left(\mathcal{O}_{q4}^{\alpha\beta}-\mathcal{O}_{q4}^{\beta\alpha}
    \right)\,.
\end{align}
In a first step both matrix elements and Wilson coefficients are divergent, which can be cured by OPE renormalization. Below, we summarize the derivation of the OPE result of Ref.~\cite{Bijnens:2022itw} and then include gluonic $\alpha_s$ corrections.

\subsection{Quark operators through $D=4$ without gluonic corrections}\label{sec:D4nogluons}
\begin{figure}
    \begin{center}
     \includegraphics[width=0.9\textwidth]{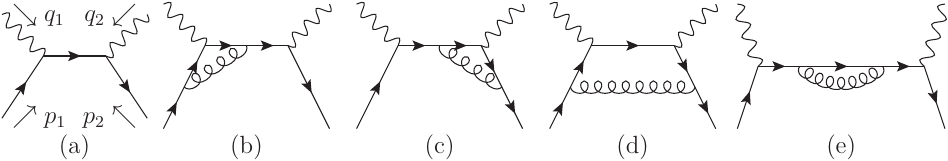}   
    \end{center}
    \caption{\label{fig:OPEquark}The diagrams relevant for the OPE of two photon currents. (a) Lowest order (b-e) Gluonic corrections. The diagrams with $q_1$ and $q_2$ interchanged need to be added.}
\end{figure}

We start by considering the OPE without gluonic corrections, order by order in $\hat{q}$. For this case, only the diagram in Fig.~\ref{fig:OPEquark}(a) is relevant\footnote{The momenta $p_1$ and $p_2$ in the diagram are assumed to be small compared to $\hat{q}$. }. Two operators appear through NLO ($D=4$), namely
\begin{align}
    \mathcal{O}_{q3}^{\alpha\beta\gamma} &= \overline q(0)\left(
    \gamma^{\alpha}\gamma^{\beta}\gamma^{\gamma}
    -\gamma^{\gamma}\gamma^{\beta}\gamma^{\alpha}\right)q(0)
    \, ,
    \nonumber\\
    \mathcal{O}_{q4}^{\alpha\beta} &= \overline q(0)\gamma^\beta\left(
    \overrightarrow D^\alpha-\overleftarrow D^\alpha\right)q(0) 
    \, . 
\end{align}
The first one is the $D=3$ contribution studied first in Ref.~\cite{Melnikov:2003xd}. The leading term in the inverse $\hat{q}$ expansion of the diagram of Fig.~\ref{fig:OPEquark}(a) and its symmetric addition with $q_1 \leftrightarrow q_2$ leads to the well-known result at $D=3$,
\begin{align}\label{eq:JJOPE}
i
& 
\int \frac{d^{4}q_4}{(2\pi)^4}\int d^{4}x_1\int d^{4}x_2\, e^{-i(q_1 x_1+q_2 x_2)}
\, 
T\{ J_j^{\mu_1}(x_1)J_{j}^{\mu_2}(x_2)\} 
  = 
  \frac{\hat q_\beta}{\hat q^2}\, g^{\mu_1}_\alpha g^{\mu_2}_\gamma\mathcal{O}_{q3}^{\alpha\beta\gamma} + \ldots \, .  
\end{align}
On the right-hand side we see the Wilson coefficient multiplying the operator, for which we have omitted the flavor index $j$.

At leading order in $\alpha_s$, we can identify the operator with the axial current, 
\begin{equation}\label{eq:axialop}
\mathcal{O}^{\alpha\beta\gamma}_{q3}\to -2i\epsilon^{\delta\alpha\beta\gamma}\bar{q}\gamma_{\delta}\gamma_5 q \, ,
\end{equation}
whose Lorentz decomposition can be written as\footnote{An overall minus sign was missed in Eqs. (5.5) and (5.6) of Ref.~\cite{Bijnens:2022itw}, without any subsequent impact in the results presented in the reference.} 
\begin{align}\nonumber
&\sum_{j}\frac{e_{q,j}^2}{e^2}\lim_{q_4 \rightarrow 0}\, i\, \partial_{q_4}^{\nu_4}\langle  0|\bar{q}_{j} \gamma_{\delta}\gamma_5 q_{j}|\gamma^{\mu_3}(q_3)\gamma^{\mu_4}(q_4)\rangle
\\
&=-
\frac{1}{4\pi^2}\, q_{3}^2\, \left(\epsilon_{\mu_3\mu_4\nu_4\delta}\, \omega_{T}(q_3^{2})
-\frac{1}{q_3^2}\, \epsilon_{q_3\mu_4\nu_4\delta}\, q_{3\mu_3}\, \omega_{T}(q_3^{2})
+\frac{1}{q_3^2}\, \epsilon_{\mu_3 \mu_4 \nu_4 q_3}\, q_{3\delta}\, \left[\omega_{L}(q_3^2)-\omega_{T}(q_3^2)\right] \right) \, .
\end{align}
Here $\epsilon ^{0123} = +1 $ and the form factors $\omega_L(q_3^2)$ and $\omega_T(q_3^2)$ are normalized according to $2\, \omega _T = \omega _L = 2/q_{3}^2$ when $q_3$ is in the perturbative limit $-q_3^2\gg \Lambda _{\textrm{QCD}}^2$.

\begin{figure}
    \begin{center}
     \includegraphics[width=0.75\textwidth]{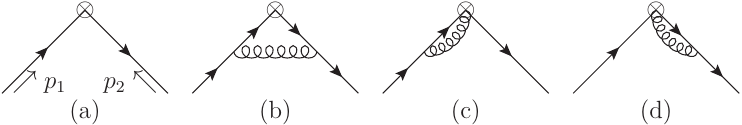}   
    \end{center}
    \caption{\label{fig:OPEquark2}The diagrams relevant for the quark matrix element of the operators, where the crossed vertex indicates the insertion of an operator. (a) Lowest order (b-d) Gluonic corrections.}
\end{figure}

The second operator $\mathcal{O}_{q4}^{\alpha\beta}$ is the $D=4$ correction~\cite{Bijnens:2022itw,Colangelo:2019lpu}. In Ref.~\cite{Bijnens:2022itw} we decomposed the quark-current matrix element $\mathcal{O}_{q4}^{\alpha\beta}$ for flavor structure $F$ according to 
\begin{equation}
\sum_j  \mathcal{T}^{(F)}_{j} 
\lim_{q_4 \rightarrow 0}\, \partial ^{\nu_4}_{q_4} \, \Big\langle 
\mathcal{O}_{q4}^{\alpha\beta}
\Big\rangle ^{j, \mu_3,\, \mu_4}=\sum_{i=1}^6 \omega_{(F)}^{D,i}\, L_i^{\alpha\beta\mu_3\mu_4\nu_4} \, . \label{eq:octetME}
\end{equation}
Here the $\omega_{(F)}^{D,i} = \omega_{(F)}^{D,i} (q_3^2)$ are non-perturbative form factors, satisfying the linear relations
\begin{align}\label{eq:kincancelrels}
    \omega _{(F)}^{D,2} & = -2\, \omega _{(F)}^{D,1}+ \omega _{(F)}^{D,5} -\frac{\omega _{(F)}^{D,6}}{2}-\frac{\omega _{T,(F)}Q_i^2}{8\pi^2 } \, ,
    \nonumber \\
    \omega _{(F)}^{D,3} & = -2\, \omega _{(F)}^{D,1}+ \omega _{(F)}^{D,5} -\frac{\omega _{(F)}^{D,6}}{2}+\frac{\omega _{T,(F)}Q_i^2}{8\pi^2 }
    \, ,
    \nonumber \\
    \omega _{(F)}^{D,4} & =\omega _{(F)}^{D,5} \, . 
\end{align}
These non-perturbative relations were derived from the cancellation of spurious kinematical singularities in the scalar functions entering the HLbL contribution, and effectively reduce the number of independent form factors to three. The Lorentz structures $L_i^{\alpha\beta\mu_3\mu_4\nu_4}$ are given by
\begin{align}\nonumber
L^{\alpha\beta\mu_3\mu_4\nu_4}_1&=g^{\mu_3\mu_4} q_3^{\nu_4} g^{\alpha\beta}-(\mu_4 \leftrightarrow \nu_4)
\, ,          \\ \nonumber
L^{\alpha\beta\mu_3\mu_4\nu_4}_2&=g^{\beta \mu_4}q_{3}^{\nu_4}\left(g^{\alpha\mu_3} -\frac{q_3^{\alpha} q_3^{\mu_3}}{q_3^2}\right)- (\mu_4 \leftrightarrow \nu_4) \, ,\\ \nonumber
L^{\alpha\beta\mu_3\mu_4\nu_4}_3&=L^{\beta\alpha\mu_3\mu_4\nu_4}_2 \, ,\\ \nonumber
L^{\alpha\beta\mu_3\mu_4\nu_4}_4&=g^{\alpha\mu_4}q_3^{\beta}\left(g^{\mu_3\nu_4} -\frac{q_3^{\mu_3} q_3^{\nu_4}}{q_3^2}\right) - (\mu_4 \leftrightarrow \nu_4) \, ,\\ \nonumber
L^{\alpha\beta\mu_3\mu_4\nu_4}_5&=L^{\beta\alpha\mu_3\mu_4\nu_4}_4 \, ,\\
L^{\alpha\beta\mu_3\mu_4\nu_4}_6&= \frac{1}{q_3^2} \left[g^{\mu_3\mu_4} q_3^{\nu_4} q^{\alpha}_3 q^{\beta}_3-(\mu_4 \leftrightarrow \nu_4)\right]   \, . 
\end{align}
$\mathcal{O}_{q4}^{\alpha\beta}$ can actually be decomposed into several operators that renormalize differently. Indeed, the symmetric traceless ($\mathcal{O}_{q4+}^{\alpha\beta}$), tracelike ($\mathcal{O}_{q4+}^{\alpha\beta}=\frac{g^{\alpha\beta}}{d}\mathcal{O}_{q4+}{}^{\gamma}_{\gamma}$) and antisymmetric ($\mathcal{O}_{q4-}^{\alpha\beta}$) in $\alpha\beta$ parts are in different Lorentz representations, where
\begin{align}\label{eq:Oq4albet}
    \mathcal{O}_{q4}^{\alpha\beta}= 
\mathcal{O}_{q4,\delta}^{\alpha\beta}+\mathcal{O}_{q4+}^{\alpha\beta}+ \mathcal{O}_{q4-}^{\alpha\beta}  .
\end{align}
We can therefore split the contributions as follows
\begin{align}
    L^{\alpha\beta\mu_3\mu_4\nu_4}_1 
    & =
    L^{\alpha\beta\mu_3\mu_4\nu_4}_{+,1}+L^{\alpha\beta\mu_3\mu_4\nu_4}_{\delta,1} 
    \, ,
    \\
     L^{\alpha\beta\mu_3\mu_4\nu_4}_2  
     & =
     L^{\alpha\beta\mu_3\mu_4\nu_4}_{+,2} + L^{\alpha\beta\mu_3\mu_4\nu_4}_{-,2} 
     \, ,
     \\
     L^{\alpha\beta\mu_3\mu_4\nu_4}_3  
     & =
     L^{\alpha\beta\mu_3\mu_4\nu_4}_{+,2} - L^{\alpha\beta\mu_3\mu_4\nu_4}_{-,2} \, ,
     \\
     L^{\alpha\beta\mu_3\mu_4\nu_4}_4
     & =
     L^{\alpha\beta\mu_3\mu_4\nu_4}_{+,4} + L^{\alpha\beta\mu_3\mu_4\nu_4}_{-,4} \, ,
     \\
     L^{\alpha\beta\mu_3\mu_4\nu_4}_5  
     & =
     L^{\alpha\beta\mu_3\mu_4\nu_4}_{+,4} - L^{\alpha\beta\mu_3\mu_4\nu_4}_{-,4} \, ,
     \\
     L^{\alpha\beta\mu_3\mu_4\nu_4}_6  
     & =-d \, 
     L^{\alpha\beta\mu_3\mu_4\nu_4}_{+,1}-2L^{\alpha\beta\mu_3\mu_4\nu_4}_{+,2}+2\,L^{\alpha\beta\mu_3\mu_4\nu_4}_{+,4} \, .
\end{align}
It then follows that the associated form factor relations between the two bases are
\begin{align}
\omega_{(F)}^{D, 1}&
=\omega_{(F),+}^{D, 1} +\omega_{(F),\delta}^{D,1}
=\omega_{(F),+}^{D, 1} 
\, ,\\
\omega_{(F)}^{D,2}&=\omega_{(F),+}^{D,2}+\omega_{(F),-}^{D,2}
=\omega_{(F),+}^{D,2}-\frac{Q^2_3 \, \omega_{T,(F)}}{8\pi^2} 
\, ,\\
\omega_{(F)}^{D,3}&=\omega_{(F),+}^{D,2}-\omega_{(F),-}^{D,2}=\omega_{(F),+}^{D,2}+\frac{Q^2_3 \, \omega_{T,(F)}}{8\pi^2} 
\, ,\\
\omega_{(F)}^{D,4}&=\omega_{(F),+}^{D,4}+\omega_{(F),-}^{D,4}=\omega_{(F),+}^{D,4} 
\, , \\
\omega_{(F)}^{D,5}&=\omega_{(F),+}^{D,4}-\omega_{(F),-}^{D,4}=\omega_{(F),+}^{D,4} 
\, , \\
\omega_{(F)}^{D,6}&
=-d\, \omega_{(F),+}^{D,1}-2\omega_{(F),+}^{D,2}+2\omega_{(F),+}^{D,4} 
\, .
\end{align}
In the last equality of each row 
we have imposed the condition of cancellation of spurious kinematic singularities of Eq.~(\ref{eq:kincancelrels}),
\begin{equation}
\omega_{(F),\delta}^{D,1}=0 \, ,\quad
\omega^{D,4}_{(F),-}=0 \, ,\quad
\omega^{D,2}_{(F),-}=-\frac{Q_3^2\omega_{T,(F)}}{8\pi^2} \, .
\end{equation}
Rewritten in this way, it can be verified that they can also be obtained by applying the quark equation of motion. Specifically, this requires that the trace-like component of the matrix element vanishes ($\omega_{(F),\delta}^{D,1}=0$) and, through Eq.(\ref{eq:eom}), that the matrix elements of the antisymmetric component are fully determined by the transverse form factor of the axial current, $\omega_{T,(F)}$. This is explicitly shown in Appendix~\ref{app:axialcurrent}.\footnote{We have explicitly verified that these conditions are satisfied for the perturbative matrix elements studied.} As we will explicitly show in Sec.~\ref{sec:amucancellations}, due to large non-trivial cancellations between terms, the $D=4$ quark matrix element only contributes with $\omega_{T,(F)}$ to the leading term of the $a_\mu ^{\textrm{HLbL}}$ integrand.

\begin{figure}
    \centering
    \includegraphics{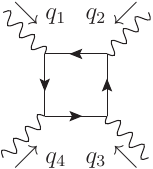}
    \caption{The perturbative quark loop. }
    \label{fig:OPEFIELDLO}
\end{figure}

Including also the contributions from the field-strength operators through the topology of Fig.~\ref{fig:OPEFIELDLO}, one finds the OPE form of the derivative tensor through $D=4$. Adding also the OPE renormalization, needed to have separately finite Wilson coefficients and matrix elements, one finds (see Ref.~\cite{Bijnens:2022itw} for details)\footnote{The notation $\langle A \rangle ^{j,\mu _3 ,\mu _4}_{\overline{\textrm{MS}}(\mu)}  =  \langle 0 | A_{j}| \gamma ^{\mu _3}(q_3) \gamma ^{\mu _4}(q_4)\rangle _{\overline{\textrm{MS}}(\mu)}$, where $j$ is a flavor index and $\overline{\textrm{MS}}(\mu)$ denotes the renormalization prescription of the matrix element. }~\cite{Bijnens:2022itw}
\begin{small}
\begin{equation}\begin{aligned}
&\lim_{q_4\rightarrow 0} \frac{\partial \Pi^{\mu_1\mu_2\mu_3\nu_4}}{\partial q_{4\, \mu_4}}
=
\sum_j\frac{ e_{q,j}^2}{e^2}\,   \lim_{q_4 \rightarrow 0}\partial ^{\nu_4}_{q_4}\, \Bigg\langle\bar{q}(0)\left[
\Gamma^{\mu_1\mu_2}(-\hat{q})-\Gamma^{\mu_2\mu_1}(-\hat{q})
\right]
q(0) \Bigg\rangle^{j, \mu_3,\mu_4} 
\\
&
+\sum_j\frac{ie_{q_j}^2}{e^2\hat{q}^2}\, 
\left(g^{\mu_1\delta}g^{\mu_2}_{\beta}+g^{\mu_2\delta}g^{\mu_1}_{\beta}-g^{\mu_1\mu_2}g^{\delta}_{\beta}\right)
\left(g_{\alpha \delta}-2\, \frac{\hat{q}_\delta \hat{q}_\alpha}{\hat{q}^2}\right)
 \\
& 
\qquad 
\times 
\lim_{q_4 \rightarrow 0}\partial_{\nu_4}^{q_4}\, \Bigg\langle \bar{q}(0)(\overrightarrow{D}^{\alpha}-\overleftarrow{D}^{\alpha})\gamma^{\beta}  q(0) \Bigg\rangle ^{j, \mu_3,\, \mu_4}_{\overline{\mathrm{MS}}(\mu)}  \\
 \\
&
+\sum_j\frac{ie_{q_j}^2}{e^2\hat{q}^2}\, 
\left(g^{\mu_1\delta}g^{\mu_2}_{\beta}+g^{\mu_2\delta}g^{\mu_1}_{\beta}-g^{\mu_1\mu_2}g^{\delta}_{\beta}\right)
\left(g_{\alpha \delta}-2\, \frac{\hat{q}_\delta \hat{q}_\alpha}{\hat{q}^2}\right) 
 \\
&
\qquad
\times 
\lim_{q_4 \rightarrow 0}\partial^{\nu_4}_{q_4} \, 
\Bigg\langle
Z^j_{DF}(\mu)\, \frac{\alpha}{4\pi}\, \left(F^{\mu\nu}F_{\mu\nu}g^{\alpha\beta}+d\, F^{\alpha\gamma}F_{\gamma}^{\, \,  \beta}\right) \label{eq:masteroperen}
 \\
& \qquad \qquad \qquad
+ Z^j_{DG}(\mu)\, \frac{\alpha_s}{4\pi}\, \left( G^{\mu\nu}_a G_{\mu\nu}^a \,  g^{\alpha\beta}+d\, G^{\alpha\gamma}_a G_{\gamma}^{a , \, \beta}\right) 
\Bigg\rangle^{j,\mu_3\mu_4}
 \\
&
+ \sum_j   \frac{ e_{q_j}^{2}}{8 e^2} \lim_{q_4 \rightarrow 0} \left[ \partial^{\nu_4}_{q_4}\, \Bigg\langle e^2 e_{q_j}^2  F_{\nu_3'\mu_3'}F_{\nu_4'\mu_4'} +\frac{1}{2N_{c}}\,  g_s^2 \,  G^{a}_{\nu_3'\mu_3'}G^{a}_ {\nu_4'\mu_4'} \Bigg\rangle^{j,\mu_3\mu_4}\right] \, 
\lim_{q_3,q_4 \rightarrow 0}\partial_{q_3}^{\nu_3'}\, \partial_{q_4}^{\nu_4'}\, \Pi_{\mathrm{ql},j}^{\mu_1\mu_2\mu_3 '\mu_4 '}
\, .
\end{aligned}
\end{equation}
\end{small}
The renormalization factors defined at scale $\mu$ in $d=4-2\epsilon$ dimensions are given by 
\begin{align}\label{eq:renfactorsqed}
Z^j_{DF}(\mu)
& =
-i\, \frac{2}{3}\,  N_c \, e_{q,j}^2\, \frac{\mu^{-2\epsilon}}{\hat{\epsilon}} \, ,
 \\
Z^j_{DG}(\mu)
& =
-i\, \frac{1}{3}\,  \frac{\mu^{-2\epsilon}}{\hat{\epsilon}} \, ,
\end{align}
and the Dirac structure $\Gamma ^{\mu_1 \mu_2}(q) = \gamma ^{\mu _1}S(q)\gamma ^{\mu _2}$, with $S(q) = \slashed{q}/q^2$ the chiral limit quark propagator. 
On the last line in the OPE a Wilson coefficient related to the perturbative quark loop appears, given by the following correlation function with all quark fields contracted
\begin{equation}\label{eq:qloopwilson}
\Pi_{\mathrm{ql},j}^{\mu_1\mu_2\mu_3'\mu_4'}=
\left[-i\int \frac{d^{4}q_{4}}{(2\pi)^{4}}\left(\prod_{i=1}^{4}\int d^{4}x_{i}\, e^{-i q_{i} x_{i}}\right)  \langle 0 | T\left(\prod_{j=1}^{4}J_j^{\mu_{j}}(x_{j})\right)|0\rangle \right]_{\mathrm{quark\,  loop}} \, .
\end{equation}

As explained in section 4.2.5 of Ref.~\cite{Bijnens:2022itw}, the matrix elements of two operators emerging at different orders in $\alpha$ in the OPE, $\mathcal{O}_{q4}^{\alpha \beta}$ and $\alpha  \mathcal{O}_{FF,i}^{\alpha \beta}$, lead to total contributions of the same order in $\alpha$ to the HLbL tensor, considering that the matrix element of the $\mathcal{O}_{q4}^{\alpha \beta}$ operator needs two extra long-distance QED vertices to capture the external photon states and then give a nonzero contribution. In this special situation, it is straightforward to check that one needs to account for the $\mathcal{O}_{q4}^{\alpha \beta} \to \alpha  \mathcal{O}_{FF,i}^{\alpha \beta}$ QED mixing when renormalization is performed to find the correct leading order result.

\begin{table}[t]\centering
\renewcommand{\arraystretch}{1.5}
\begin{tabular}{|c|c|c|c|}
\hline
& $\overline{Q}_1$ & $\overline{Q}_2$ & $\overline{Q}_3$
 \\
 \hline 
$\hat{\Pi}_1$  & 5 & 5 & 4 \\ \hline
$\hat{\Pi}_4$  & 3 & 3 & 5 \\ \hline
$\hat{\Pi}_7$  & 4 & 5 & 6 \\ \hline
$\hat{\Pi}_{17}$ & 5 & 5 & 5 \\ \hline
$\hat{\Pi}_{39}$ & 5 & 5 & 5 \\ \hline
$\hat{\Pi}_{54}$ & 5 & 5 & 5
\\ \hline
\end{tabular}
\caption{\label{tab:truncorder}The leading power $k$ of $1/ \overline{Q}_i^{k}$ where the studied OPE through $D=4$ has no prediction. }
\renewcommand{\arraystretch}{1}
\end{table}

To obtain the contribution from the above OPE to $a_\mu^{\textrm{HLbL}}$ one first needs to project the derivative tensor to the six scalar functions $\hat{\Pi}_{1,4,7,17,39,54}$, as thoroughly described in Ref.~\cite{Bijnens:2022itw}. The orders the OPE here cannot predict are given in Table~\ref{tab:truncorder}. The scalar functions through the predictive powers are given by
\begin{align}
\underline{q_{3}\textrm{ small:}} &
\nonumber \\
\hat{\Pi}_1&=\frac{2}{\pi^2 \overline{Q}_3^2}\omega_{L}(q_3^2) 
\, ,\\
\hat{\Pi}_4&=-\frac{64\left( \omega_{+}^{D,1}-\omega_{+}^{D,4}\right)  }{\overline{Q}_{3}^4} 
\, ,\\
\hat{\Pi}_7
&
=
\mathcal{O}\left( \frac{1}{\overline{Q}_{3}^6}\right)
\, ,\\
\hat{\Pi}_{17}
&=\frac{32 \left(-3 (\omega_{+}^{D,1}-\omega_{+}^{D,4})-(\omega_{+}^{D,1}+\omega_{+}^{D,2})-\frac{\omega_{T}Q_3^2}{8\pi^2}\right) }{\overline{Q}_{3}^4 Q_{3}^2} \, ,
\\
\hat{\Pi}_{39}&=-\frac{32 \left(- (\omega_{+}^{D,1}-\omega_{+}^{D,4})+(\omega_{+}^{D,1}+\omega_{+}^{D,2})+\frac{\omega_{T}Q_3^2}{8\pi^2}\right) }{\overline{Q}_{3}^4 Q_{3}^2} \, ,
\\
\hat{\Pi}_{54}&=\mathcal{O}\left( \frac{1}{\overline{Q}_{3}^5}\right) \, ,
\end{align}

\begin{align}
\underline{q_{1}\textrm{ small:}} &
\nonumber \\
\hat{\Pi}_1&=\frac{32\, \omega_{+}^{D,4} }{\overline{Q}_{1}^4} \, ,\\
\hat{\Pi}_4&=\frac{8 \left(- (\omega_{+}^{D,1}-\omega_{+}^{D,4})+(\omega_{+}^{D,1}+\omega_{+}^{D,2})+\frac{\omega_{T}Q_1^2}{8\pi^2}\right) }{\overline{Q}_{1}^2 Q_{1}^2}\, ,\\
\hat{\Pi}_7&=\mathcal{O}\left(\frac{1}{\overline{Q}_{1}^4} \right) \, ,\\
\hat{\Pi}_{17}&=\mathcal{O}\left(\frac{1}{\overline{Q}_{1}^5} \right) \, ,\\
\hat{\Pi}_{39}&=-\frac{32 \left(- (\omega_{+}^{D,1}-\omega_{+}^{D,4})+(\omega_{+}^{D,1}+\omega_{+}^{D,2})+\frac{\omega_{T}Q_1^2}{8\pi^2}\right) }{\overline{Q}_{1}^4 Q_{1}^2} ,\\
\hat{\Pi}_{54}&=\frac{32 \left(- (\omega_{+}^{D,1}-\omega_{+}^{D,4})+(\omega_{+}^{D,1}+\omega_{+}^{D,2})-\frac{\omega_{T}Q_1^2}{8\pi^2}\right) }{\overline{Q}_{1}^4 Q_{1}^2} \, ,
\end{align}

\begin{align}
\underline{q_{2}\textrm{ small:}} & \nonumber \\
\hat{\Pi}_1&=\frac{32}{\overline{Q}_{2}^{4}}\, \omega_{+}^{D,4} \, ,\\
\hat{\Pi}_4&=\frac{8 \left(- (\omega_{+}^{D,1}-\omega_{+}^{D,4})+(\omega_{+}^{D,1}+\omega_{+}^{D,2})+\frac{\omega_{T}Q_2^2}{8\pi^2}\right) }{\overline{Q}_{2}^2 Q_{2}^2}\,  ,
\\
\hat{\Pi}_7&=4\frac{8 \left(- (\omega_{+}^{D,1}-\omega_{+}^{D,4})+(\omega_{+}^{D,1}+\omega_{+}^{D,2})+\frac{\omega_{T}Q_2^2}{8\pi^2}\right) }{\overline{Q}_{2}^4 Q_{2}^2} \, , \\
\hat{\Pi}_{17}&=\mathcal{O}\left(\frac{1}{\overline{Q}_{2}^5} \right) \, ,\\
\hat{\Pi}_{39}&=-4\frac{8 \left(- (\omega_{+}^{D,1}-\omega_{+}^{D,4})+(\omega_{+}^{D,1}+\omega_{+}^{D,2})+\frac{\omega_{T}Q_2^2}{8\pi^2}\right) }{\overline{Q}_{2}^{4} Q_{2}^2}  \, ,\\
\hat{\Pi}_{54}&=-4\frac{8  \left(- (\omega_{+}^{D,1}-\omega_{+}^{D,4})+(\omega_{+}^{D,1}+\omega_{+}^{D,2})-\frac{\omega_{T}Q_2^2}{8\pi^2}\right) }{\overline{Q}_{2}^4 Q_{2}^2} \, ,
\end{align}
with the $\omega^{D,i}_{+}$ defined just above. The extra photon field contributions from the last four lines in Eq.~(\ref{eq:masteroperen}) are
\begin{align}\label{eq:pihatff}
\underline{q_{3}\textrm{ small:}} &
\nonumber \\
\hat{\Pi}_4 & =-\frac{16}{3\pi^2 \overline{Q}_3^4} \, ,   
\nonumber \\
\underline{q_{1}\textrm{ small:}} &
\nonumber \\
 \hat{\Pi}_1& =\frac{8}{3\pi^2 \overline{Q}_1^4}\left(1-4\log\frac{\overline{Q}_1}{2\mu}\right) \, ,
\nonumber \\
\underline{q_{2}\textrm{ small:}} &
\nonumber \\
 \hat{\Pi}_1 & =\frac{8}{3\pi^2 \overline{Q}_2^4}\left(1-4\log\frac{\overline{Q}_2}{2\mu}\right) \, .   
\end{align}
At this order in the perturbative regime
\begin{align}\label{eq:omipertLO1}
\omega^{(+)}&\equiv \omega_{+}^{D,1}+\omega_{+}^{D,2}= -\frac{1}{24\pi^2} \, ,\\
\omega^{(-)}&\equiv \omega_{+}^{D,1}-\omega_{+}^{D,4}= 0   \, ,
\label{eq:omipertLO2}\\
\omega_{+}^{D,4}&=\frac{-13+12\log \frac{Q_i}{\mu}}{36\pi^2} \, ,
\label{eq:omipertLO3}
\end{align}
where everything is in $\sum_q N_{c}e_{q}^4$ units for the sum of the flavor octet and singlet pieces. Indeed, these reproduce the results from the overlap region to the OPE studied in Ref.~\cite{Bijnens:2021jqo}, where the scale $\mu$ cancels to yield only logarithms of ratios of the high and low scales. As an explicit example, in the perturbative regime,
\begin{equation}\label{eq:pertexample}
\hat{\Pi}_1^{q_1}=\underbrace{\frac{8}{3\pi^2 \overline{Q}_1^4}\left(1-4\log\frac{\overline{Q}_1}{2\mu}\right)}_{ \mathcal{O}_{\{FF\}}}+\underbrace{\frac{32}{\overline{Q}_{1}^4} \frac{-13+12\log \frac{Q_1}{\mu}}{36\pi^2}}_{\mathcal{O}_{q4}}=\frac{16}{\overline{Q}_1^4}\frac{-5-6\log\frac{\overline{Q}_1}{2Q_1}}{9\pi^2} \, ,
\end{equation}
which exactly matches the expanded quark loop result of Ref.~\cite{Bijnens:2021jqo}. In the underbraces we have indicated where the separate pieces come from, suppressing Lorentz indices and indicating the presence of several operators involving $F$ as $\{FF\}$ (see the last three lines of Eq.~(\ref{eq:masteroperen})). 

\subsection{Perturbative gluonic corrections through $D=4$}\label{sec:gluonic}
In this section we consider novel gluonic order-$\alpha _s$ corrections to the OPE through dimension $D=4$.
For these corrections one needs to calculate the diagrams in Fig.~\ref{fig:OPEquark}, and in order to obtain the Wilson coefficients\footnote{Quark wave function renormalization diagrams are not needed, they drop out of the Wilson coefficients.} we also need to determine the operator matrix elements from the diagrams shown in Fig.~\ref{fig:OPEquark2}. It should be noted that there are gluonic corrections to both the perturbative Wilson coefficients as well as the operator matrix elements. Below we separately consider $D=3$ and $D=4$ in the OPE. 

\subsubsection{Gluonic corrections at $D=3$}
We now return to Eq.~(\ref{eq:JJOPE}) and consider the gluonic corrections from the diagrams in Fig.~\ref{fig:OPEquark}(b-e), working with dimensional regularization. The $D=3$ result is finite
\begin{align}
i %\antonio{\cancel{\sum_{j,k}\frac{ e_{q_j}e_{q_k}}{e^{2}}}}
& 
\int \frac{d^{4}q_4}{(2\pi)^4}\int d^{4}x_1\int d^{4}x_2\, e^{-i(q_1 x_1+q_2 x_2)}
\, 
T\{ J_j^{\mu_1}(x_1)J_j^{\mu_2}(x_2)\} 
  \nonumber \\
  &
  = 
  \left(1-\frac{7}{3}\frac{\alpha_s}{\pi}\right)  \, \frac{\hat q_\beta}{\hat q^2}g^{\mu_1}_\alpha g^{\mu_2}_\gamma\mathcal{O}_{q3,0}^{\alpha\beta\gamma} + \ldots
 \, .  
\end{align}
In the perturbative regime, we can compute the gluonic correction to the matrix element in dimensional regularization, finding a finite result
\begin{equation}
\langle \mathcal{O}_{q3,0}^{\alpha\beta\gamma}\rangle_{\alpha_s}= \frac{4}{3}\alpha_s\langle \mathcal{O}_{q3,0}^{\alpha\beta\gamma}\rangle_{\mathrm{quark\, loop}} \, .
\end{equation}
This is not incompatible with the non-renormalization theorems requiring the lack of $\alpha_s$ corrections for the axial current that respects chiral symmetry. The identification of Eq.~(\ref{eq:axialop}), 
\begin{equation}
\langle j_{5\delta}^{0} \rangle= \frac{-i}{12}\, \epsilon_{\delta\alpha\beta\gamma} \langle \mathcal{O}^{\alpha\beta\gamma}_{q3,0} \rangle \, ,
\end{equation}
where the matrix element on the right-hand side has been computed at $d$ dimensions, is equivalent to the axial current with HV prescription for regulating $\gamma_5$~\cite{Trueman:1979en}, which breaks chiral symmetry. In fact, it is related with the `genuine' axial current through a finite renormalization~\cite{Trueman:1979en}
\begin{equation}
j_5^0=Z_{A}^{-1}\, j_5=\left(1+\frac{4}{3}\frac{\alpha_s}{\pi}\right) j_5 \, \to \mathcal{O}^{\alpha\beta\gamma}_{q3,0}=Z_A^{-1} \, \mathcal{O}^{\alpha\beta\gamma}_{q3} \, .
\end{equation}
Indeed the matrix element corresponding to the $\mathcal{O}^{\alpha\beta\gamma}_{q3}$ does not receive any $\alpha_s$ correction and one finds
\begin{align}
i 
& 
\int \frac{d^{4}q_4}{(2\pi)^4}\int d^{4}x_1\int d^{4}x_2\, e^{-i(q_1 x_1+q_2 x_2)}
\, 
T\{ J_j^{\mu_1}(x_1)J_j^{\mu_2}(x_2)\} 
  \nonumber \\
  &
  = 
  \left(1-\frac{\alpha_s}{\pi}\right) \frac{\hat q_\beta}{\hat q^2}g^{\mu_1}_\alpha g^{\mu_2}_\gamma\mathcal{O}_{q3}^{\alpha\beta\gamma}  + \ldots
  \, .  
\end{align}
We thus see that the Wilson coefficient for the operator associated to the axial current that respects chiral symmetry indeed is corrected as $\left(1-\frac{\alpha_s}{\pi}\right)$, which was conjectured in Ref.~\cite{Ludtke:2020moa}. In fact, this result can also be obtained from our previous work~\cite{Bijnens:2021jqo} in the limit of perturbative corner kinematics $\Lambda _{\textrm{QCD}}^2 \ll Q_3^2\ll Q_{1}^2,Q_{2}^2$, when considering the functions $\tilde{\Pi}_i$ and $\hat{\Pi}_i$ (which can be obtained through projection techniques as described in Ref.~\cite{Bijnens:2022itw}). For the derivative tensor, we thus have
\begin{align}
    \lim _{q_4\to 0}\frac{\partial \Pi ^{\mu _1 \mu_2 \mu_3 \nu_4}}{\partial q_{4\, \mu _4}} 
     & 
     \stackrel{D=3}{=} 
      \left(1-\frac{\alpha_s}{\pi}\right) \, \left. \lim _{q_4\to 0} \frac{\partial \Pi ^{\mu _1 \mu_2 \mu_3 \nu_4}}{\partial q_{4\, \mu _4}}
     \right|_{\alpha _{s} = 0} \, ,
\end{align}
where the derivative tensor evaluated at $\alpha _s = 0$ on the right-hand side is given by the $D=3$ contribution in Eq.~(\ref{eq:masteroperen}).

\subsubsection{Gluonic corrections at $D=4$}
\begin{figure}
    \centering
    \includegraphics[width=0.8\textwidth]{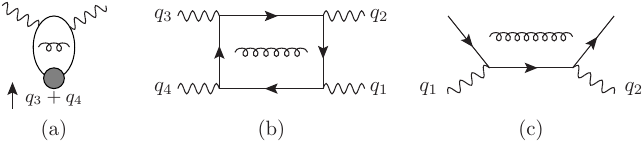}
    \caption{Diagrams needed for the gluonic corrections to the OPE through $D=4$. The gluon line represents all possible contractions, and the dark vertex in (a) an insertion of the relevant quark operator. It should be noted that (c) corresponds to the set of diagrams in Fig.~\ref{fig:OPEquark}(a)-(e).}
    \label{fig:gluondiagrams}
\end{figure}

We next introduce gluonic corrections at dimension $D=4$, first in the bare case and later we renormalize the obtained OPE. To simplify this technical calculation and clarify how the various pieces contribute, we introduce new simplified notation. This will be particularly convenient for the renormalization procedure carried out below and also unify the whole OPE with and without gluonic corrections, leading to the central result of Eq.~(\ref{eq:gluonicrenope}). In the new notation, the full non-renormalized OPE at $D=4$ can be written\footnote{In the following we will suppress the Lorentz indices $\mu_1$, $\mu_2$, $ \mu _3$, $\mu _4$ and $\nu _4$ as they do not play a central role. As will be seen, the only relevant indices here are $\alpha$ and $\beta$ as the parts symmetric and antisymmetric renormalize differently. It is important to note that the matrix elements in this section implicitly contain the derivative $\partial _{q_4}$ evaluated in the static limit $q_4\rightarrow 0$, as in e.g.~Eq.~(\ref{eq:masteroperen}). }
\begin{align}
      \lim _{q_4\to 0} \frac{\partial \Pi ^{\mu _1 \mu_2 \mu_3 \nu_4}}{\partial q_{4\, \mu _4}}      = 
    C_{4,\alpha \beta}
     \, \langle 
    \mathcal{O}_{q4}^{\alpha \beta}\rangle
    + C_{F}\langle \mathcal{O}_{F}\rangle 
     +  \tilde{C}_{4,\alpha \beta}\, \langle \tilde{\mathcal{O}}_{q4}^{\alpha \beta}\rangle  \, .
     \end{align}
The quantities on the right-hand side include gluon corrections to both Wilson coefficients and matrix elements, and through an expansion in $\alpha _s$ we will obtain the OPE studied above as well as the novel gluonic corrections. It should be understood that the quantities on the right-hand side are bare, non-renormalized Wilson coefficients and operators. We here see that the operator $\tilde{\mathcal{O}}_{q4}^{\alpha\beta}$ from Eq.~(\ref{eq:redundantop}) appears at NLO in $\alpha_s$, which by virtue of the equation of motion can be related to $\mathcal{O}_{q4}^{\alpha \beta}$ as in Eq.~(\ref{eq:eom}), multiplied by the Wilson coefficient $\tilde{C}_{4,\alpha \beta}$. In the absence of gluonic corrections, the matrix element $\langle \mathcal{O}_{F} \rangle$ is the one appearing on the last line of Eq.~(\ref{eq:masteroperen}) and $C_{F}$ the Wilson coefficient corresponding to the perturbative quark loop multiplying it. To first order in $\alpha_s$ the matrix elements and Wilson coefficients can be expanded according to 
\begin{align}
\langle \mathcal{O}_{q4}^{\alpha \beta} \rangle &=\langle 
    \mathcal{O}_{q4}^{\alpha \beta}(0,1) \rangle+ \langle \mathcal{O}_{q4}^{\alpha \beta}(1,2) \rangle \, ,\\
    C_{4}^{\alpha \beta} &= C_{4,\alpha \beta}(0,0) + C_{4,\alpha \beta}(1,0) \, , \\
    \tilde{C}_{4}^{\alpha \beta} &=  \tilde{C}_{4}^{\alpha \beta}(1,0) \, ,\\
    C_{F}&=C_F(0,1)+C_F(1,2) \, ,
\end{align}
so that the HLbL tensor is given by\footnote{Explicit expressions for the bare Wilson coefficients and perturbative matrix elements are provided in the supplementary material file {\tt WilsonPertME.txt}.}
\begin{align}\label{eq:startope_nonren}
     \frac{\partial \Pi ^{\mu _1 \mu_2 \mu_3 \nu_4}}{\partial q_{4\, \mu _4}} 
     & 
     = 
    C_{4,\alpha \beta}(0,0)
     \, \langle 
    \mathcal{O}_{q4}^{\alpha \beta}(0,1) \rangle
    + C_{F}(0,1)\langle \mathcal{O}_{F}(0,0)\rangle 
    \nonumber \\
    & 
    + C_{4,\alpha \beta}(0,0)\, \langle \mathcal{O}_{q4}^{\alpha \beta}(1,2) \rangle
    + C_{4,\alpha \beta}(1,1)
    \,  \langle \mathcal{O}_{q4}^{\alpha \beta}(0,1) \rangle 
    \nonumber
    \\
    &
     +  \tilde{C}_{4,\alpha \beta}(1,0)\, \langle \tilde{\mathcal{O}}_{q4}^{\alpha \beta}(0,0)\rangle 
     + C_{F}(1,2)\, \langle \mathcal{O}_{F}(0,0) \rangle
    \, .
\end{align}
Here $\langle \mathcal{O}_{q4}^{\alpha \beta}(a,b) \rangle$, $\langle \tilde{\mathcal{O}}_{q4}^{\alpha \beta}(a,b) \rangle$ and $\langle \mathcal{O}_{F}(a,b) \rangle$ are the indicated matrix elements at order $\alpha _s^a$ with a leading divergence in $\epsilon $ as $1/\epsilon^{b}$, i.e.
\begin{align}
    \langle \mathcal{O}^{\alpha \beta}(a,b) \rangle 
     \sim \langle \tilde{\mathcal{O}}^{\alpha \beta}(a,b) \rangle \sim 
    \alpha _s^a \, \left(\frac{1}{\epsilon ^b} +\dots \right) \, . 
\end{align}
With this notation in place, $\langle \mathcal{O}_{F}(0,0) \rangle$ and $\langle \tilde{\mathcal{O}}^{\alpha \beta}(0,0) \rangle$ appearing for the derivative tensor are finite. The quark matrix elements $ \langle \mathcal{O}^{\alpha \beta}(a,b) \rangle$ and $ \langle \tilde{\mathcal{O}}^{\alpha \beta}(a,b) \rangle$ are diagrammatically represented in Fig.~\ref{fig:gluondiagrams}(a), where the relevant operator is inserted at the vertex and gluons should be included in the diagram if $a=1$. Again, the matrix element $\langle \mathcal{O}_{F}(0,0) \rangle$ is the one appearing on the last line of Eq.~(\ref{eq:masteroperen}) in terms of the electromagnetic field-strength tensor\footnote{Here we work in the perturbative limit, where the additional matrix element contributions from the gluon field-strength tensor $G_{\mu \nu}^a$ (only present for the singlet parts) do not contribute at the studied order.} $F_{\mu \nu}$. 

The Wilson coefficients $C_{q4,\alpha \beta}(a,b)$ and $\tilde{C}_{q4,\alpha \beta}(a,b)$ for given $a$ and $b$ are analogously to above defined in terms of the scaling in $\epsilon$ and $\alpha _s$. They are associated to the diagrams in Fig.~\ref{fig:gluondiagrams}(c), where the gluons should be included if $a=1$. The Wilson coefficients $C_{F}(a,b)$ are associated to the topology of Fig.~\ref{fig:gluondiagrams}(b), either with ($C_{F}(1,2)$) or without the gluon line ($C_{F}(0,1)$).

With the above notation, it follows that the first line in Eq.~(\ref{eq:startope_nonren}) is the non-re-normalized OPE at $D=4$ without gluonic corrections, first obtained in Ref.~\cite{Bijnens:2022itw}. The last two lines in Eq.~(\ref{eq:startope_nonren}) enter at order $\alpha _s$ and are the new additions of this work. We emphasise that Eq.~(\ref{eq:startope_nonren}) is finite as $\epsilon \to 0$, but not renormalized since the individual Wilson coefficients and matrix elements are not (cf.~Eq.~(\ref{eq:masteroperen})). Before discussing renormalization, we elaborate on the various contributions appearing above.  

\paragraph{Wilson coefficients:} Here we consider $C_{4,\alpha \beta} (0,0)$, $C_{4,\alpha \beta} (1,1)$ and  $\tilde{C}_{4,\alpha \beta} (1,0)$. The first contribution, $C_{4,\alpha \beta} (0,0)$ is obtained from the diagram in Fig.~\ref{fig:gluondiagrams}(c) without gluons. Concretely, one expands the diagram at NLO in powers of $\frac{p_{1,2}}{\hat{q}}$ (where again $p_{1,2}$ are the momenta on the quark legs) and identify the new emerging spinor structures as the ones coming from the corresponding $D=4$ operator. The factor multiplying that spinor structure can then be identified as $C_{4,\alpha \beta} (0,0)$. Similarly, 
the diagrams with the gluonic corrections in Fig.~\ref{fig:gluondiagrams}(c), after taking $p_{1,2}$ much smaller than both $\hat{q}$ and the loop momentum, lead to the $D=4$ Wilson coefficients $C_{4,\alpha \beta} (1,1)$ and $\tilde{C}_{4,\alpha \beta} (1,0)$. 

 The coefficient $C_{F}(1,2)$ is related to the gluon-corrected quark loop in Fig~\ref{fig:gluondiagrams}(b), through
\begin{align}
 C_{F}(1,2) = \lim _{q_3 , q_4 \to 0} \partial _{q_3}^{\nu_3 '}\partial _{q_4}^{\nu_4 '} \Pi ^{\mu _1 \mu _2 \mu_3 ' \mu_4 '}_{\textrm{ql,gluon}}\, , 
\end{align}
where $\Pi ^{\mu _1 \mu _2 \mu_3 \mu_4}_{\textrm{ql,gluon}}$ is the $\mathcal{O}(\alpha _s)$ corrected quark loop in Fig~\ref{fig:gluondiagrams}(b). The primed indices here are contracted against the matrix element $\langle \mathcal{O}_F (0,0)\rangle$ as in Eq.~(\ref{eq:masteroperen}). 

At fixed external momenta $q_3$ and $q_4$ there are 10 possible diagrams in Fig.~\ref{fig:gluondiagrams}(b), of which 4 are self-energy corrections to the quark propagators and 6 that connect different quark propagators. Taking into account permutations of the external momenta this yields 30 different diagrams. To obtain $C_{F}(1,2)$ we first take the derivatives in $q_3$ and $q_4$ and take the soft limit $q_3,q_4 \rightarrow 0$. This results in a set of 2-loop tensor integrals to calculate. To handle these we Lorentz decompose in the external momentum $\hat{q}$, which is the only external scale in the diagram, such that only scalar loop integrals remain. The scalar integrals are of the form
\begin{align}
    I_{a,b,c,d,e}(\hat{q}) = \frac{1}{i^2} \int \frac{d^dp_1}{(2\pi)^d}\frac{d^d p_2}{(2\pi)^4} \frac{1}{p_1^{2a} \, (p_1-\hat{q})^{2b}\, p_2^{2c} \, (p_2-\hat{q})^{2d} \, (p_1-p_2)^{2e} } \, .
\end{align}
Using {\tt Kira}~\cite{Maierhoefer:2017hyi}, the integrals from the diagrams can be reduced to two master integrals,
\begin{align}
I_{1,1,1,1,0}& = B_1^2 \, ,
\\
I_{0,1,1,0,1}& = S_1 \, ,
\end{align}
which respectively correspond to the standard squared 1-loop bubble integral and the 2-loop sunset integral. These divergent integrals result in the leading $1/\epsilon^2$ in $C_{F}(1,2)$. The two master integrals are discussed in detail in Ref.~\cite{Bijnens:2021jqo} and references therein, and we therefore leave out further details here.

\paragraph{Quark matrix elements:}
For $\langle \mathcal{O}_{q4}^{\alpha \beta}(0,1) \rangle$, $\langle \mathcal{O}_{q4}^{\alpha \beta}(1,2) \rangle$ and $\langle \tilde{\mathcal{O}}_{q4}^{\alpha \beta}(0,0)\rangle$ one has to calculate the diagrams in Fig.~\ref{fig:gluondiagrams}(a) with the respective operators inserted in the bottom vertex. In our reduced notation the quark operator matrix elements contain an external derivative in $q_4^{\nu_4}$ that should be applied to the diagram before sending $q_4 \rightarrow 0$. For instance, in the case of $\langle \mathcal{O}_{q4}^{\alpha \beta}(0,1) \rangle$ the operator $\mathcal{O}_{q4}^{\alpha \beta}$ is inserted at the bottom vertex and all contractions performed, after which $\lim _{q_4 \to 0}\partial _{q_4}^{\nu_4}$ is applied. 

For each of $\langle \mathcal{O}_{q4}^{\alpha \beta}(0,1) \rangle$ and $\langle \tilde{\mathcal{O}}_{q4}^{\alpha \beta}(0,0)\rangle$ there are two 1-loop diagrams appearing, corresponding to the swap of $q_3$ and $q_4$ on the external legs. For $\langle \mathcal{O}_{q4}^{\alpha \beta}(1,2) \rangle$, on the other hand, one has to add all possible ways the gluon can connect to both the quark legs as well as the operator. This yields 9 2-loop diagrams to calculate, of which 3 are self-energy type diagrams, 3 connect different quark legs and 3 connect the operator to the quark legs. For the matrix elements considered here, a reduction to scalar master integrals yields only the bubble $B_1$ and the sunset $S_1$. 

\paragraph{Renormalization:} Having calculated the bare matrix elements and Wilson coefficients, one finds that the sum of contributions in Eq~(\ref{eq:startope_nonren}) is finite. To renormalize the OPE, including the new corrections, we next perform the necessary steps to arrive at the final renormalized OPE through $D=4$ in Eq.~(\ref{eq:gluonicrenope}). As discussed in section~\ref{sec:D4nogluons}, it is convenient to separate the terms into antisymmetric and symmetric tensors in $\alpha$ and $\beta$. This separation allows Eq.~(\ref{eq:startope_nonren}) to split into two distinct parts.
\begin{align}\label{eq:startope_nonren_pm}
     \lim _{q_4\to 0} \frac{\partial \Pi ^{\mu _1 \mu_2 \mu_3 \nu_4}}{\partial q_{4\, \mu _4}} 
   & = 
    \lim _{q_4\to 0} \left( \frac{\partial \Pi ^{\mu _1 \mu_2 \mu_3 \nu_4}}{\partial q_{4\, \mu _4}} \right) _{-}
    +   \lim _{q_4\to 0}\left( \frac{\partial \Pi ^{\mu _1 \mu_2 \mu_3 \nu_4}}{\partial q_{4\, \mu _4}} \right) _{+}
    \, ,
    \\
    \lim _{q_4\to 0}  \left( \frac{\partial \Pi ^{\mu _1 \mu_2 \mu_3 \nu_4}}{\partial q_{4\, \mu _4}} \right) _{-}
     & 
     =  C_{4,\alpha \beta}^{-}
     \, \langle 
    \mathcal{O}_{q4,-}^{\alpha \beta} \rangle 
   +  \tilde{C}_{4,\alpha \beta}^{-}\, \langle \tilde{\mathcal{O}}_{q4,-}^{\alpha \beta}\rangle \, ,
    \\
     \lim _{q_4\to 0} \left( \frac{\partial \Pi ^{\mu _1 \mu_2 \mu_3 \nu_4}}{\partial q_{4\, \mu _4}} \right) _{+}
     & 
     =   C_{4,\alpha \beta}^{+}
     \, \langle 
    \mathcal{O}_{q4,+}^{\alpha \beta}\rangle
    + C_{F}\langle \mathcal{O}_{F}\rangle 
    \, .
\end{align}
The two terms in the antisymmetric part are separately finite, whereas the symmetric contribution contains intermediate divergences that cancel between terms and therefore needs to be renormalized. Below we separately study the antisymmetric and symmetric derivative tensors in more detail.

The antisymmetric piece can be written on a simpler form by exploiting the quark equation of motion dictating that
\begin{equation}
\langle \tilde{\mathcal{O}}_{q4,-}^{\alpha \beta}\rangle  
     =
    4 \langle \mathcal{O}_{q4,-}^{\alpha \beta}\rangle \, .
\end{equation}
We have explicitly checked this holds in the perturbative regime discussed earlier. Moreover, one finds $\tilde{C}_{4,\alpha \beta}^{-}(1,0)=\frac{5}{8}\, C_{4,\alpha \beta}^{-}(1,0)$ which means that the whole antisymmetric contribution can be expressed as
\begin{equation}
  \lim _{q_4\to 0} \left( \frac{\partial \Pi ^{\mu _1 \mu_2 \mu_3 \nu_4}}{\partial q_{4\, \mu _4}} \right) _{-} 
     = 
      \bar{C}_{4,\alpha \beta}^{-} 
 \langle 
    \mathcal{O}_{q4,-}^{\alpha \beta} \rangle 
    \, .  
\end{equation}
where 
\begin{equation}
\bar{C}_{4,\alpha \beta}^{-}=\left[ C_{4,\alpha \beta}^{-}(0,0) +\frac{7}{2}\, C_{4,\alpha \beta}^{-}(1,0)\right] \, . 
\end{equation}

For the symmetric piece, we first make the observation that $\langle \mathcal{O}_{q4,+}^{\alpha \beta}\rangle$ is traceless. This implies that only the traceless part $\hat{C}_{4,\alpha \beta}^{+}$ of $C_{4,\alpha \beta}^{+}$ contributes to the derivative tensor,
\begin{align}
    C_{4,\alpha \beta}^{+}
     \, \langle 
    \mathcal{O}_{q4,+}^{\alpha \beta}\rangle = \hat{C}_{4,\alpha \beta}^{+}
     \, \langle 
    \mathcal{O}_{q4,+}^{\alpha \beta}\rangle \, .
\end{align}
We next introduce the renormalization scale $\mu$, such that the symmetric bare quark operator $\mathcal{O}_{q4,+}^{\alpha \beta}$ is renormalized through 
\begin{align}
\mathcal{O}_{q4,+}^{\alpha \beta}=[1+\delta Z_{q4}(\mu )]\, \mathcal{O}_{q4,+}^{\alpha \beta}(\mu ) +[Z_{F}(\mu) + \delta Z_{F}(\mu) ]\, \alpha\, \mathcal{O}_{FF,+}^{\alpha \beta} 
\, .
\end{align}
Here $\delta Z_{q4}(\mu )$ and $\delta Z_{F}(\mu)$ are renormalization factors starting at order $\alpha_{s}$, $Z_{F}(\mu)$ is at order $\alpha_s ^0$ and the finite $\mathcal{O}_{FF,+}^{\alpha \beta} = (d\, \mathcal{O}_{FF,1,+}^{\alpha \beta}+\mathcal{O}_{FF,2,+}^{\alpha \beta})$ is independent of $\mu$ at the studied order. Neglecting the QCD corrections gives back the renormalization in Ref.~\cite{Bijnens:2022itw}, associated to the mixing of $ \alpha \mathcal{O}_{FF,+}^{\alpha \beta} $ and $\mathcal{O}_{q4,+}^{\alpha \beta}$. Specifically, expanding the above expression yields
\begin{align}
   \langle \mathcal{O}_{q4,+}^{\alpha \beta}(0,1) \rangle 
   &= 
   \langle \mathcal{O}_{q4,+}^{\alpha \beta}(0,0) \rangle _{\overline{\textrm{MS}}(\mu)} + Z_{F}(\mu ,0,1) \,  \langle \alpha\,  \mathcal{O}_{FF,+}^{\alpha \beta}(0,0) \rangle \, , 
   \\
    \langle \mathcal{O}_{q4,+}^{\alpha \beta}(1,2) \rangle 
    & =
    \langle \mathcal{O}_{q4,+}^{\alpha \beta}(1,0) \rangle _{\overline{\textrm{MS}}(\mu)} 
    + \delta Z_{q4}(\mu , 1,1) \, \langle \mathcal{O}_{q4,+}^{\alpha \beta}(0,0) \rangle_{\overline{\textrm{MS}}(\mu)}
  \nonumber \\
  & + \delta Z_{F}(\mu ,1,2)\, \langle \alpha\, 
 \mathcal{O}_{FF,+}^{\alpha \beta}(0,0) \rangle 
   \nonumber \\
  & =
   \langle \mathcal{O}_{q4,+}^{\alpha \beta}(1,0) \rangle _{\overline{\textrm{MS}}(\mu)} 
   +  \delta Z_{q4}(\mu , 1,1) \,  \langle \mathcal{O}_{q4,+}^{\alpha \beta}(0,1) \rangle 
   \nonumber \\
   &
    - \delta Z_{q4}(\mu , 1,1) \,  Z_{F}(\mu ,0,1) \,  \langle \alpha\, \mathcal{O}_{FF,+}^{\alpha \beta}(0,0) \rangle
   + \delta Z_{F}(\mu ,1,2)\, \langle \alpha\, 
 \mathcal{O}_{FF,+}^{\alpha \beta}(0,0) \rangle
    \, . 
\end{align}
In the last equality we reinserted the renormalization condition for $\langle \mathcal{O}_{q4,+}^{\alpha \beta}(0,1) \rangle$ to relate $ \langle \mathcal{O}_{q4,+}^{\alpha \beta}(1,2) \rangle $ to its renormalized counterpart and bare quantities. The renormalization constants are found to be\footnote{It should be noted that $Z_{F}(\mu,0,1)$ of course is directly related to $Z_{DF}^{j}(\mu)$ in Eq.~(\ref{eq:renfactorsqed}), but has here been rescaled to fit the notation.}
\begin{align}
Z_{F}(\mu,0,1) & = -i\, \frac{2\, d_{G}}{3}\, \frac{\mu^{-2\epsilon}}{\hat{\epsilon}} \, ,
  \, \\
  \delta Z_{q4}(\mu,1,1) 
  & =   -\frac{8\, d_G}{3} \, \frac{\mu ^{-2\epsilon}}{\hat{\epsilon}} \, \alpha _s
  \, ,
  \\
   \delta Z_{F}(\mu,1,2) 
  & =    
  i\, d_G^2 \left[ \frac{8}{9}\, \frac{\mu^{-4\epsilon}}{\hat{\epsilon} ^2} - \frac{49}{27}\, \frac{\mu^{-2\epsilon}}{\hat{\epsilon} }\right]\, \alpha _s
  \, ,
\end{align}
with $d_G = (4\pi)^\epsilon /(16\pi^2)\, \Gamma (1+\epsilon)\Gamma (1-\epsilon)^2/\Gamma (1-2\epsilon)$ which for $\epsilon \to 0$ is just $d_G = 1/(16\pi^2)$. The OPE for the symmetric tensor can thus be written on the renormalized form
\begin{align}
    \lim _{q_4\to 0} \left( \frac{\partial \Pi ^{\mu _1 \mu_2 \mu_3 \nu_4}}{\partial q_{4\, \mu _4}} \right) _{+}
     & 
     =  \left\{  \hat{C}_{4,\alpha \beta}^{+} \left[ 1+\delta Z_{q4}(\mu )\right] \right\} 
     \, \langle 
    \mathcal{O}_{q4,+}^{\alpha \beta}\rangle _{\overline{\textrm{MS}}(\mu)} 
    \nonumber \\
    &
    + \left\{  \hat{C}_{4,\alpha \beta}^{+} \left[Z_{F}(\mu) + \delta Z_{F}(\mu) \right]\right\} \, \langle \alpha\, \mathcal{O}_{FF,+}^{\alpha \beta} \rangle 
   + C_{F}\,  \langle \mathcal{O}_{F}\rangle \, . 
\end{align}
The first and second rows are separately finite. In expanded form the same OPE reads 
\begin{align}
   \lim _{q_4\to 0} \left( \frac{\partial \Pi ^{\mu _1 \mu_2 \mu_3 \nu_4}}{\partial q_{4\, \mu _4}} \right) _{+}
     & 
     =  \left\{  
     \hat{C}_{4,\alpha \beta}^{+}(0,0)
     +\hat{C}_{4,\alpha \beta}^{+}(0,0)\, \delta Z_{q4}(\mu,1,1 )
     + \hat{C}_{4,\alpha \beta}^{+}(1,1)
      \right\} 
      \nonumber \\
      & 
      \times
     \left\{\langle 
    \mathcal{O}_{q4,+}^{\alpha \beta}(0,0)\rangle _{\overline{\textrm{MS}}(\mu)} +\langle 
    \mathcal{O}_{q4,+}^{\alpha \beta}(1,0)\rangle _{\overline{\textrm{MS}}(\mu)}\right\}
    \nonumber \\
    &
    + \Big\{  
     \hat{C}_{4,\alpha \beta}^{+}(0,0)\, Z_{F}(\mu,0,1)
     +\hat{C}_{4,\alpha \beta}^{+}(0,0)\, \delta Z_{F}(\mu ,1,2)
    \nonumber \\
    &
      + \hat{C}_{4,\alpha \beta}^{+}(1,1)\, Z_{F}(\mu,0,1)
      \Big\} 
      \times 
       \langle \alpha\, \mathcal{O}_{FF,+}^{\alpha \beta} (0,0)\rangle 
   \nonumber \\
      &
       + \left\{C_{F}(0,1)+C_{F}(1,2)\right\}\langle \mathcal{O}_{F}(0,0)\rangle \, . 
\end{align}

The total renormalized OPE for the symmetric and antisymmetric pieces added together with the $D=3$ contribution of the last section is
\begin{align}\label{eq:gluonicrenope}
     \lim _{q_4\to 0} \frac{\partial \Pi ^{\mu _1 \mu_2 \mu_3 \nu_4}}{\partial q_{4\, \mu _4}} 
     & = 
     \left(1-\frac{\alpha_s}{\pi}\right) \, \left. \lim _{q_4\to 0} \frac{\partial \Pi ^{\mu _1 \mu_2 \mu_3 \nu_4}}{\partial q_{4\, \mu _4}}
     \right|_{\alpha _{s} = 0}^{D=3}
     \nonumber \\
      & +\bar{C}_{4,\alpha \beta}^{-} 
 \langle 
    \mathcal{O}_{q4,-}^{\alpha \beta} \rangle 
    + \left\{  \hat{C}_{4,\alpha \beta}^{+} \left[ 1+\delta Z_{q4}(\mu )\right] \right\} 
     \, \langle 
    \mathcal{O}_{q4,+}^{\alpha \beta}\rangle _{\overline{\textrm{MS}}(\mu)} 
    \nonumber \\
    &
    + \left\{  \hat{C}_{4,\alpha \beta}^{+} \left[Z_{F}(\mu) + \delta Z_{F}(\mu) \right]\right\} \, \langle \alpha\, 
 \mathcal{O}_{FF,+}^{\alpha \beta} \rangle 
   + C_{F}\,  \langle \mathcal{O}_{F}\rangle \, . 
\end{align}

 Here we may perturbatively collect everything to order $\alpha _s$. This means that the same matrix element decomposition into $\overline{\omega} _{+}^{D,i}$, $\omega_L$ and $\omega _{T}$ holds, but each form factor can be expanded in $\alpha _ s$. We write this expansion as
\begin{align}\label{eq:omegaalphas}
 \overline{\omega} _{+}^{D,i} = \omega _{+}^{D,i} + \alpha _s \, \delta \omega _{+}^{D,i} \, ,
\end{align}
where $\omega _{+}^{D,i}$ are the order $\alpha _s^0$ contributions satisfying Eqs.~(\ref{eq:omipertLO1})--(\ref{eq:omipertLO3}) in the perturbative limit. The $D=3$ contribution is given in terms of $\omega _L$ and $\omega _T$. With this in place we can extract the $\hat{\Pi}_i$ associated to the gluonic corrections through $D=4$. In units of $N_c \sum e_q^4$ the results in the three regions are \footnote{These expressions are given in the supplementary file {\tt RenPiHatsD4Q123.txt}.}
\begin{align}
\underline{q_{3}\textrm{ small:}} &
\nonumber \\
\hat{\Pi}_1&\stackrel{\alpha_s}{=} 
-\alpha _s\, \frac{2    \, \omega _L }{ \pi^3\, \overline{Q}_3^2}
\, ,\\
\hat{\Pi}_4&\stackrel{\alpha_s}{=}   -\alpha _s\, \frac{64 \Big[8 \, \delta _{12}^2\, 
   (\omega _{+}^{D,1}-\omega _{+}^{D,4})+9 \, 
   (\delta \omega _{+}^{D,1}-\delta \omega _{+}^{D,4}) \, \pi  Q_3^2\Big]}{9 \pi \, 
   Q_3^2 \overline{Q}_3^4}
   \nonumber \\
   & 
   +\alpha _s\, \frac{64 \Big[ 16  \log \frac{\overline{Q}_3}{2\mu} \,  (\omega _{+}^{D,1}-
   \omega _{+}^{D,4})+(\omega _{+}^{D,1}+\omega _{+}^{D,4})\Big]}{9 \pi 
  \overline{Q}_3^4}
\, ,\\
\hat{\Pi}_7
&
\stackrel{\alpha_s}{=}
\alpha _s \frac{512\,  \delta _{12}\,  (\omega _{+}^{D,2}+\omega _{+}^{D,4})}{9 \pi\,  
   Q_3^2 \overline{Q}_3^5}
\, ,\\
\hat{\Pi}_{17}
&\stackrel{\alpha_s}{=}
\alpha _s \, \frac{32 \Big[ (16\log \frac{\overline{Q}_3}{2\mu}-1) (4\, 
   \omega _{+}^{D,1}+\omega _{+}^{D,2}-3 \, \omega _{+}^{D,4})-9 (4\, 
   \delta \omega _{+}^{D,1}+\delta \omega _{+}^{D,2}-3 \, \delta \omega _{+}^{D,4}) \pi \Big]}{9 \pi  \, Q_3^2 \overline{Q}_3^4}
   \nonumber \\
   &
 +\alpha _s\,  \frac{4\,   \omega_T
   }{ \pi ^3  \overline{Q}_3^4}
\, ,
\\
\hat{\Pi}_{39}&\stackrel{\alpha_s}{=} \alpha _s \frac{  32\Big[  (16\log \frac{\overline{Q}_3}{2\mu}-1)
   (\omega _{+}^{D,2}+\omega _{+}^{D,4})-9 (\delta \omega _{+}^{D,2}+\delta \omega _{+}^{D,4})
   \pi \Big]}{9 \pi \,  Q_3^2 \overline{Q}_3^4}
   \nonumber \\
   &
   +\alpha _s \frac{4 \,  \omega _T }{ \pi ^3 \,  \overline{Q}_3^4}
\, ,\\
\hat{\Pi}_{54}&\stackrel{\alpha_s}{=} \mathcal{O}\left( \frac{\alpha _s}{\overline{Q}_3^5}\right) 
\, ,
\end{align}
\begin{align}
\underline{q_{1}\textrm{ small:}} &
\nonumber \\
\hat{\Pi}_1&\stackrel{\alpha_s}{=} 
\alpha _s \, \frac{32  \, \Big[9 \, \delta \omega _{+}^{D,4} \, \pi + \omega _{+}^{D,4}( -16 \log \frac{\overline{Q}_1}{2\mu}\, 
   +1)\Big]}{9 \pi\,   \overline{Q}_1^4}
\, ,\\
\hat{\Pi}_4&\stackrel{\alpha_s}{=}
\alpha _s \,
\frac{ 8 \Big[9 (\delta \omega _{+}^{D,2}+\delta \omega _{+}^{D,4}) \pi
   -(16 \log \frac{\overline{Q}_1}{2\mu} -1) (\omega _{+}^{D,2}+\omega _{+}^{D,4})\Big] }{9 \pi \,   Q_1^2 \overline{Q}_1^2}
   \nonumber \\
   & 
  -\alpha _s \,
\frac{  \omega _T
   }{ \pi ^3  \overline{Q}_1^2}
\, ,\\
\hat{\Pi}_7&\stackrel{\alpha_s}{=} 
\mathcal{O}\left( \frac{\alpha _s}{\overline{Q}_1^4}\right) 
\, ,\\
\hat{\Pi}_{17}&\stackrel{\alpha_s}{=} 
\mathcal{O}\left( \frac{\alpha _s}{\overline{Q}_1^5}\right) 
\, ,\\
\hat{\Pi}_{39}&\stackrel{\alpha_s}{=} 
\alpha _s \, \frac{32  \Big[ (16 \log \frac{\overline{Q}_1}{2\mu}-1)
   (\omega _{+}^{D,2}+\omega _{+}^{D,4})-9 (\delta \omega _{+}^{D,2}+\delta \omega _{+}^{D,4})
   \pi \Big]}{9 \pi \,  Q_1^2 \overline{Q}_1^4}
   \nonumber \\
& +\alpha _s \, \frac{4 \, \omega _T }{ \pi ^3\,  \overline{Q}_1^4}   
\, ,\\
\hat{\Pi}_{54}&\stackrel{\alpha_s}{=} 
\alpha _s \, \frac{32 \Big[9 (\delta \omega _{+}^{D,2}+\delta \omega _{+}^{D,4})
   \pi -(16\log \frac{\overline{Q}_1}{2\mu}-1) (\omega _{+}^{D,2}+\omega _{+}^{D,4})\Big]}{9 \pi \,  Q_1^2 \overline{Q}_1^4}
      \nonumber \\
& +\alpha _s \, \frac{4 \, \omega _T }{ \pi ^3\,  \overline{Q}_1^4}
\, ,
\end{align}
\begin{align}
\underline{q_{2}\textrm{ small:}} &
\nonumber \\
\hat{\Pi}_1&\stackrel{\alpha_s}{=}
\alpha _s \, \frac{32 \Big[9 \delta \omega _{+}^{D,4} \pi + \omega _{+}^{D,4} (-16 \log \frac{\overline{Q}_2}{2\mu}
   +1)\Big]
   }{9 \pi  \overline{Q}_2^4}
\, ,\\
\hat{\Pi}_4&\stackrel{\alpha_s}{=}
\alpha _s \,
\frac{ 8  \Big[9 (\delta \omega _{+}^{D,2}+\delta \omega _{+}^{D,4}) \pi
   -(16 \log \frac{\overline{Q}_2}{2\mu}-1) (\omega _{+}^{D,2}+\omega _{+}^{D,4})\Big] }{9 \pi \,  Q_2^2 \overline{Q}_2^2}
 \nonumber \\
 & - \alpha _s \,
\frac{\omega _T
   }{ \pi ^3 \,  \overline{Q}_2^2}
\, ,\\
\hat{\Pi}_7&\stackrel{\alpha_s}{=}
 -  \alpha _s \, 
 \frac{32 \Big[(16 \log \frac{\overline{Q}_2}{2\mu}-1)
   (\omega _{+}^{D,2}+\omega _{+}^{D,4})-9 (\delta \omega _{+}^{D,2}+\delta \omega _{+}^{D,4})
   \pi \Big]}{9 \pi \,   Q_2^2 \overline{Q}_2^4}
   \nonumber \\
   & 
   -  \alpha _s \, 
 \frac{4 \, \omega _T }{ \pi ^3  \overline{Q}_2^4}
\, ,\\
\hat{\Pi}_{17}&\stackrel{\alpha_s}{=}
\mathcal{O}\left( \frac{\alpha _s}{\overline{Q}_2^5}\right) 
\, ,\\
\hat{\Pi}_{39}&\stackrel{\alpha_s}{=}
\alpha _s \,
\frac{32 \Big[ (16 \log \frac{\overline{Q}_2}{2\mu}-1)
   (\omega _{+}^{D,2}+\omega _{+}^{D,4})-9 (\delta \omega _{+}^{D,2}+\delta \omega _{+}^{D,4})
   \pi \Big]}{9 \pi  Q_2^2 \overline{Q}_2^4}
   \nonumber \\
   &
   + 
   \alpha _s \,
\frac{4 \, \omega _T }{ \pi ^3 \,  \overline{Q}_2^4}
\, ,\\
\hat{\Pi}_{54}&\stackrel{\alpha_s}{=}
-\alpha _s \, \frac{32 \Big[ 9 (\delta \omega _{+}^{D,2}+\delta \omega _{+}^{D,4})
   \pi -(16 \log \frac{\overline{Q}_2}{2\mu}-1) (\omega _{+}^{D,2}+\omega _{+}^{D,4})\Big]}{9 \pi \, Q_2^2 \overline{Q}_2^4}
   \nonumber \\
   &
   -\alpha _s \, \frac{4 \,  \omega _T
   }{ \pi ^3\,   \overline{Q}_2^4}
\, .
\end{align}
Again, in the perturbative corner kinematics limit the above result agrees with our Ref.~\cite{Bijnens:2021jqo} through the predictive powers of the OPE (see Table~\ref{tab:truncorder}). The matching relations are\footnote{We emphasize that beyond the perturbative regime the expansion of Eq.~(\ref{eq:omegaalphas}) does not make sense. In that regime, one simply adds  the $\hat{\Pi}_i$ $\alpha_s$ corrections of this section without the $\delta \omega_+^{D,4}$ factors to the results of the $\hat{\Pi}_i$ in Sec.~(\ref{sec:D4nogluons}) in terms of $\overline{\omega} _{+}^{D,i}$.}
\begin{align}
\delta \omega _{+}^{D,2} + \delta \omega _{+}^{D,4} & = -\frac{5+8\, \log \frac{Q_i}{\mu}}{108\pi ^3}
\, ,
\\
\delta \omega _{+}^{D,4}-\delta \omega _{+}^{D,1} & =  -\frac{7}{108\pi^3}
\, , 
\\
\delta \omega _{+}^{D,4} & = \frac{11035 - 36\, \log \frac{Q_i}{\mu} + 288\,  \log ^2 \frac{Q_i}{\mu } + 432 \, S_1(2)}{972 \pi^3} \, .
\end{align}
Here $S_1(2) = -865/32+3\, \zeta _{3}/2$. 

For the last line of the renormalized OPE in Eq.~(\ref{eq:gluonicrenope}), associated to the field-strength terms, we find the $\hat{\Pi}_i$ functions at order $\alpha _s$ to be
\begin{align}
 \underline{q_{3}\textrm{ small:}} &
\nonumber \\
\hat{\Pi}_4 & \stackrel{\alpha_s}{=} -\alpha _s\, \frac{64  \left(-22 -54\,  \zeta _3+30 \log \frac{\overline{Q}_3}{2\mu}\right)}{405 \pi ^3\, 
   \overline{Q}_3^4}\, ,   
\nonumber \\
\underline{q_{1}\textrm{ small:}} &
\nonumber \\
 \hat{\Pi}_1& \stackrel{\alpha_s}{=}-\alpha _s\, \frac{ 4\left(-6029+2592 \, \zeta _3-2880 \log ^2 \frac{\overline{Q}_1}{2\mu}+6240
   \log  \frac{\overline{Q}_1}{2\mu} \right)}{1215 \pi ^3 \, \overline{Q}_1^4} \, ,
\nonumber \\
\underline{q_{2}\textrm{ small:}} &
\nonumber \\
 \hat{\Pi}_1 & \stackrel{\alpha_s}{=} -\alpha _s\, \frac{4\left(-6029+2592\,  \zeta _3-2880 \log^2  \frac{\overline{Q}_2}{2\mu}+6240
   \log \frac{\overline{Q}_2}{2\mu}\right)}{1215 \pi ^3 \, \overline{Q}_2^4}\, .
\end{align}
These extend Eq.~(\ref{eq:pihatff}). Analogous to the case studied without gluons, adding together all $\hat{\Pi}_i$ above cancels the scale $\mu$ to yield only logarithms of ratios of the high and low momentum scales, thereby reproducing the result from Ref.~\cite{Bijnens:2021jqo}. The example of Eq.~(\ref{eq:pertexample}) in the perturbative regime generalizes, schematically, to
\begin{equation}
\begin{aligned}
\hat{\Pi}_1^{q_1}&=\underbrace{\frac{8}{3\pi^2 \overline{Q}_1^4}\left(1-4\log\frac{\overline{Q}_1}{2\mu}\right)}_{\mathcal{C}_{\{FF\}}^{\mathrm{LO}}\langle \mathcal{O}_{\{FF\}}\rangle}-\underbrace{\alpha _s\, \frac{ 4\left(-6029+2592 \, \zeta _3-2880 \log ^2 \frac{\overline{Q}_1}{2\mu}+6240
   \log  \frac{\overline{Q}_1}{2\mu} \right)}{1215 \pi ^3 \, \overline{Q}_1^4}}_{  \mathcal{C}_{\{FF\}}^{\mathrm{NLO}}\langle \mathcal{O}_{\{FF\}}\rangle}
\\
&+\underbrace{\frac{32}{\overline{Q}_{1}^4} \frac{-13+12\log \frac{Q_1}{\mu}}{36\pi^2}}_{C_{q4}^{\mathrm{LO}}\langle \mathcal{O}_{q4}^{\mathrm{LO}}\rangle}+\underbrace{\frac{32}{\overline{Q}_{1}^4}\alpha_s\frac{11035 - 36\, \log \frac{Q_i}{\mu} + 288\,  \log ^2 \frac{Q_i}{\mu } + 432 \, S_1(2)}{972 \pi^3}}_{C_{q4}^{\mathrm{LO}}\langle \mathcal{O}_{q4}^{\mathrm{NLO}}\rangle}
\\
+&\underbrace{32\alpha _s \, \frac{1 -16 \log \frac{\overline{Q}_1}{2\mu}\, 
   }{9 \pi\,   \overline{Q}_1^4}\frac{-13+12\log \frac{Q_1}{\mu}}{36\pi^2}}_{C_{q4}^{\mathrm{NLO}}\langle \mathcal{O}_{q4}^{\mathrm{LO}}\rangle}
   \\&=\frac{16}{\overline{Q}_1^4}\frac{-5-6\log\frac{\overline{Q}_1}{2Q_1}}{9\pi^2}+8\frac{\alpha_s}{405\pi^3}\left(-131+648\, \zeta_3+480\log^2\frac{\overline{Q}_1}{2Q_1}\right) \, ,
\end{aligned}
\end{equation}
which once again exactly matches the expanded two-loop result of Ref.~\cite{Bijnens:2021jqo}. In the underbraces we have indicated where each piece comes from with respect to the renormalised OPE in Eq.~(\ref{eq:gluonicrenope}), suppressing Lorentz indices and indicating the presence of several operators including $F$ as $\{FF\}$. The superscript LO refers order $\alpha_s^0$ and NLO $\alpha_s$.

\section{OPE contributions in $a_\mu^{\textrm{HLbL}}$}\label{sec:amucancellations}
In this section we investigate contributions of the derived OPE to $a_\mu^{\textrm{HLbL}}$. 

\subsection{Expanded weights}

We have applied an OPE to the derivative HLbL tensor in the $g-2$ kinematics, i.e. to
\begin{equation}
\lim_{q_4\rightarrow 0} \frac{\partial \Pi^{\mu_1\mu_2\mu_3\nu_4}}{\partial q_{4\, \mu_4}} \, ,
\end{equation}
giving a systematic expansion in negative powers of the Euclidean variables $\overline{Q}_i$, inherited by the different
scalar functions $\tilde{\Pi}_n$ and $\hat{\Pi}_n$. Taking into account that the scalar function entering into the $a_{\mu}^{\mathrm{HLbL}}$ integral, $\overline{\Pi}_{n}$, can be expressed as a linear combinations of the $\tilde{\Pi}_n$, one can systematically extract up to which order in the $1/\overline{Q}_i$ expansion one can unambiguously predict each $\overline{\Pi}_{n}$. This was done in Ref.~\cite{Bijnens:2022itw} for the directly related (through crossing) $\hat{\Pi}_n$ functions.

Let us first translate the results of this power counting exercise into the $\overline{\Pi}_n$, also using that, with the exception of $\overline{\Pi}_{1}^{q_3}$, all contributions in the M-V OPE region start from $D=4$. Taking this into account, and with that exception, we rescale the $\overline{\Pi}_{n}^{q_j}$ so that they start at most (assuming a non-zero contribution from $D=4$) at order $\mathcal{O}(1/\overline{Q}_i^0)$. This is done in Table~\ref{tab:expandedPibar}.

\begin{table}[tbh]
    \centering
    \begin{tabular}{ccc|ccc|ccc}
        \toprule
        \multicolumn{3}{c|}{\textbf{Case \( q_3 \) Small}} & 
        \multicolumn{3}{c|}{\textbf{Case \( q_1 \) Small}} & 
        \multicolumn{3}{c}{\textbf{Case \( q_2 \) Small}} \\
        \cmidrule(lr){1-3} \cmidrule(lr){4-6} \cmidrule(lr){7-9}
        \(\overline{\Pi}_i\) & & Expression & 
        \(\overline{\Pi}_i\) & & Expression & 
        \(\overline{\Pi}_i\) & & Expression \\
        \midrule
        \(\overline{\Pi}_1\) & & \(\frac{\hat{\Pi}_1^{q_3,\mathrm{R}}}{\overline{Q_3}^2}\) & 
        \(\overline{\Pi}_1\) & & \(\frac{\hat{\Pi}_1^{q_1,\mathrm{R}}}{\overline{Q_1}^4}\) & 
        \(\overline{\Pi}_1\) & & \(\frac{\hat{\Pi}_1^{q_2,\mathrm{R}}}{\overline{Q_2}^4}\) \\
        
        \(\overline{\Pi}_2\) & & \(\frac{\hat{\Pi}_1^{q_1,\mathrm{R}}}{\overline{Q_3}^4}\) & 
        \(\overline{\Pi}_2\) & & \(\frac{\hat{\Pi}_1^{q_2,\mathrm{R}}}{\overline{Q_1}^4}\) & 
        \(\overline{\Pi}_2\) & & \(\frac{\hat{\Pi}_1^{q_3,\mathrm{R}}}{\overline{Q_2}^2}\) \\
        
        \(\overline{\Pi}_3\) & & \(\frac{\hat{\Pi}_4^{q_3,\mathrm{R}}}{\overline{Q_3}^4}\) & 
        \(\overline{\Pi}_3\) & & \(\frac{\hat{\Pi}_4^{q_1,\mathrm{R}}}{\overline{Q_1}^2}\) & 
        \(\overline{\Pi}_3\) & & \(\frac{\hat{\Pi}_4^{q_2,\mathrm{R}}}{\overline{Q_2}^2}\) \\
        
        \(\overline{\Pi}_4\) & & \(\frac{\hat{\Pi}_4^{q_1,\mathrm{R}}}{\overline{Q_3}^2}\) & 
        \(\overline{\Pi}_4\) & & \(\frac{\hat{\Pi}_4^{q_2,\mathrm{R}}}{\overline{Q_1}^2}\) & 
        \(\overline{\Pi}_4\) & & \(\frac{\hat{\Pi}_4^{q_3,\mathrm{R}}}{\overline{Q_2}^4}\) \\
        
        \(\overline{\Pi}_5\) & & \(\frac{\hat{\Pi}_7^{q_3,\mathrm{R}}}{\overline{Q_3}^5}\) & 
        \(\overline{\Pi}_5\) & & \(\frac{\hat{\Pi}_7^{q_1,\mathrm{R}}}{\overline{Q_1}^3}\) & 
        \(\overline{\Pi}_5\) & & \(\frac{\hat{\Pi}_7^{q_2,\mathrm{R}}}{\overline{Q_2}^4}\) \\
        
        \(\overline{\Pi}_6\) & & \(\frac{\hat{\Pi}_7^{q_1,\mathrm{R}}}{\overline{Q_3}^3}\) & 
        \(\overline{\Pi}_6\) & & \(\frac{\hat{\Pi}_7^{q_2,\mathrm{R}}}{\overline{Q_1}^4}\) & 
        \(\overline{\Pi}_6\) & & \(\frac{\hat{\Pi}_7^{q_3,\mathrm{R}}}{\overline{Q_2}^5}\) \\
        
        \(\overline{\Pi}_7\) & & \(\frac{\hat{\Pi}_7^{q_2,\mathrm{R}}}{\overline{Q_3}^4}\) & 
        \(\overline{\Pi}_7\) & & \(\frac{\hat{\Pi}_7^{q_1,\mathrm{R}}}{\overline{Q_1}^3}\) & 
        \(\overline{\Pi}_7\) & & \(\frac{\hat{\Pi}_7^{q_3,\mathrm{R}}}{\overline{Q_2}^5}\) \\
        
        \(\overline{\Pi}_8\) & & \(\frac{\hat{\Pi}_{17}^{q_2,\mathrm{R}}}{\overline{Q_3}^4}\) & 
        \(\overline{\Pi}_8\) & & \(\frac{\hat{\Pi}_{17}^{q_3,\mathrm{R}}}{\overline{Q_1}^4}\) & 
        \(\overline{\Pi}_8\) & & \(\frac{\hat{\Pi}_{17}^{q_1,\mathrm{R}}}{\overline{Q_2}^4}\) \\
        
        \(\overline{\Pi}_9\) & & \(\frac{\hat{\Pi}_{17}^{q_3,\mathrm{R}}}{\overline{Q_3}^4}\) & 
        \(\overline{\Pi}_9\) & & \(\frac{\hat{\Pi}_{17}^{q_1,\mathrm{R}}}{\overline{Q_1}^4}\) & 
        \(\overline{\Pi}_9\) & & \(\frac{\hat{\Pi}_{17}^{q_2,\mathrm{R}}}{\overline{Q_2}^4}\) \\
        
        \(\overline{\Pi}_{10}\) & & \(\frac{\hat{\Pi}_{39}^{q_3,\mathrm{R}}}{\overline{Q_3}^4}\) & 
        \(\overline{\Pi}_{10}\) & & \(\frac{\hat{\Pi}_{39}^{q_1,\mathrm{R}}}{\overline{Q_1}^4}\) & 
        \(\overline{\Pi}_{10}\) & & \(\frac{\hat{\Pi}_{39}^{q_2,\mathrm{R}}}{\overline{Q_2}^4}\) \\
        
        \(\overline{\Pi}_{11}\) & & \(\frac{\hat{\Pi}_{54}^{q_1,\mathrm{R}}}{\overline{Q_3}^4}\) & 
        \(\overline{\Pi}_{11}\) & & \(\frac{\hat{\Pi}_{54}^{q_2,\mathrm{R}}}{\overline{Q_1}^4}\) & 
        \(\overline{\Pi}_{11}\) & & \(\frac{\hat{\Pi}_{54}^{q_3,\mathrm{R}}}{\overline{Q_2}^4}\) \\
        
        \(\overline{\Pi}_{12}\) & & \(\frac{\hat{\Pi}_{54}^{q_3,\mathrm{R}}}{\overline{Q_3}^4}\) & 
        \(\overline{\Pi}_{12}\) & & \(\frac{\hat{\Pi}_{54}^{q_1,\mathrm{R}}}{\overline{Q_1}^4}\) & 
        \(\overline{\Pi}_{12}\) & & \(\frac{\hat{\Pi}_{54}^{q_2,\mathrm{R}}}{\overline{Q_2}^4}\) \\
        \bottomrule
    \end{tabular}
    \caption{Expressions for \( \overline{\Pi}_i^{q_x} \) based on the small parameter \( q_x \). They are re-scaled in such a way that the potential $D=4$ ($D=3$ for $\hat{\Pi}_1^{q_3,\mathrm{R}}$) contributions of $\hat{\Pi}_k^{q_l}$ to $\overline{\Pi}_{i}$ are order $\mathcal{O}(1/\overline{Q}_x^0)$. Trivial cyclic permutations need to be applied in the variables whenever $l\neq x$ except for $\overline{\Pi}_{7}^{q_{1,2}}$, where one also needs to flip $y_{jk}\to -y_{jk}$.}
    \label{tab:expandedPibar}
\end{table}

Now we want to know which $\overline{\Pi}_{n}$ are actually more relevant for the $a_\mu^{\mathrm{HLbL}}$, for which we also need to expand the weights, $T_n$. Using the results of the table, let us thus do a combined $T_n\overline{\Pi}_n$ expansion in negative $\overline{Q}_3$ powers, keeping only potential contributions up to $T_n\overline{\Pi}_n\sim 1/\overline{Q}_i^6$. No contributions enter for $q_3$ small at this order.\footnote{Notice how the original crossing of the integrand inherited from the crossing of the tensor is lost due to the reshuffling of the contributions made in Refs.~\cite{Colangelo:2015ama,Colangelo:2017fiz}. The crossing under cyclic permutations can be restored by simply taking $a_\mu^{\mathrm{HLbL}}=\frac{1}{3}\sum_i a_\mu^{\mathrm{HLbL},i}$ and relabeling dummy integration variables or explicitly symmetrizing $\sum_i T_i\overline\Pi_i$ in $Q_1,Q_2,Q_3$.} Defining
\begin{equation}
\sigma_i\equiv \sqrt{1+\frac{4 m_\mu^2}{Q_i^2}} \, ,
\end{equation}
for $q_1$ small one has

% T1 \overline{\Pi}_1
\begin{equation}
\dfrac{T_1 \overline{\Pi}_1}{\hat{\Pi}_1^{q_1,\mathrm{R}}} = 0 \, ,
\end{equation}

% T2 \overline{\Pi}_2
\begin{equation}
\dfrac{T_2 \overline{\Pi}_2}{\hat{\Pi}_1^{q_1,\mathrm{R}}} = 0 \, ,
\end{equation}

% T3 \overline{\Pi}_3
\begin{equation}
\begin{aligned}
\dfrac{T_3 \overline{\Pi}_3}{\hat{\Pi}_4^{q_1,\mathrm{R}}} &= -\dfrac{8}{3 m_{\mu}^2 Q_1^2 \overline{Q}_1^6} \Bigg( -12 m_{\mu}^4 - 5 Q_1^4 ( -1 + \sigma_1 ) + 2 m_{\mu}^2 Q_1^2 ( 3 + 2 \sigma_1 ) \\
&\quad + \overline{Q}_1 \left[ 6 m_{\mu}^2 Q_1 (3 - 2 \sigma_1 ) - 3 Q_1^3 ( -1 + \sigma_1 ) \right] y_{23} \\
&\quad + 8 \left[ 6 m_{\mu}^4 + m_{\mu}^2 Q_1^2 ( -3 + \sigma_1 ) + Q_1^4 ( -1 + \sigma_1 ) \right] y_{23}^2 \Bigg) \, ,
\end{aligned}
\end{equation}

% T4 \overline{\Pi}_4
\begin{equation}
\begin{aligned}
\dfrac{T_4 \overline{\Pi}_4}{\hat{\Pi}_4^{q_2,\mathrm{R}}} &= \dfrac{16}{3 m_{\mu}^2 Q_1^2 \overline{Q}_1^6} \Bigg( -12 m_{\mu}^4 - 9 Q_1^4 ( -1 + \sigma_1 ) + 6 m_{\mu}^2 Q_1^2 (1 + 2 \sigma_1 ) \\
&\quad + 3 Q_1 \overline{Q}_1 \left[ m_{\mu}^2 (2 - 4 \sigma_1 ) + Q_1^2 ( -1 + \sigma_1 ) \right] y_{23} \\
&\quad + \left[ 48 m_{\mu}^4 + 10 Q_1^4 ( -1 + \sigma_1 ) - 4 m_{\mu}^2 Q_1^2 (3 + 2 \sigma_1 ) \right] y_{23}^2 \\
&\quad + \overline{Q}_1 \left[ 12 m_{\mu}^2 Q_1 - 6 Q_1^3 ( -1 + \sigma_1 ) \right] y_{23}^3 \\
&\quad + \left[ 8 m_{\mu}^2 Q_1^2 (3 - 2 \sigma_1 ) - 4 Q_1^4 ( -1 + \sigma_1 ) \right] y_{23}^4 \Bigg) \, ,
\end{aligned}
\end{equation}

% T5 \overline{\Pi}_5
\begin{equation}
\dfrac{T_5 \overline{\Pi}_5}{\hat{\Pi}_7^{q_1,\mathrm{R}}} = -\dfrac{16 Q_1}{3 m_{\mu}^2 \overline{Q}_1^6} \left( Q_1^2 ( -1 + \sigma_1 ) + m_{\mu}^2 ( -6 + 4 \sigma_1 ) \right) y_{23} \left( 2 + y_{23}^2 \right) \, ,
\end{equation}

% T6 \overline{\Pi}_6
\begin{equation}
\begin{aligned}
\dfrac{T_6 \overline{\Pi}_6}{\hat{\Pi}_7^{q_2,\mathrm{R}}} &= \dfrac{4}{3 m_{\mu}^2 Q_1^2 \overline{Q}_1^6} \Bigg( 12 m_{\mu}^4 + 2 m_{\mu}^2 Q_1^2 (3 - 10 \sigma_1 ) + 7 Q_1^4 ( -1 + \sigma_1 ) \\
&\quad + \overline{Q}_1 \left[ -3 Q_1^3 ( -1 + \sigma_1 ) + 6 m_{\mu}^2 Q_1 ( -1 + 2 \sigma_1 ) \right] y_{23} \\
&\quad + \left[ -48 m_{\mu}^4 + 4 m_{\mu}^2 Q_1^2 (9 - 2 \sigma_1 ) - 14 Q_1^4 ( -1 + \sigma_1 ) \right] y_{23}^2 \\
&\quad + \overline{Q}_1 \left[ -12 m_{\mu}^2 Q_1 + 6 Q_1^3 ( -1 + \sigma_1 ) \right] y_{23}^3 \\
&\quad + \left[ 4 Q_1^4 ( -1 + \sigma_1 ) + 8 m_{\mu}^2 Q_1^2 ( -3 + 2 \sigma_1 ) \right] y_{23}^4 \Bigg) \, ,
\end{aligned}
\end{equation}

% T7 \overline{\Pi}_7
\begin{equation}
\dfrac{T_7 \overline{\Pi}_7}{\hat{\Pi}_7^{q_1,\mathrm{R}}} = \dfrac{16 Q_1}{3 m_{\mu}^2 \overline{Q}_1^6} \left( Q_1^2 ( -1 + \sigma_1 ) + m_{\mu}^2 ( -6 + 4 \sigma_1 ) \right) y_{23} \left( 2 + y_{23}^2 \right) \, ,
\end{equation}

% T8 \overline{\Pi}_8
\begin{equation}
\dfrac{T_8 \overline{\Pi}_8}{\hat{\Pi}_{17}^{q_3,\mathrm{R}}} = 0 \, ,
\end{equation}

% T9 \overline{\Pi}_9
\begin{equation}
\begin{aligned}
\dfrac{T_9 \overline{\Pi}_9}{\hat{\Pi}_{17}^{q_1,\mathrm{R}}} &= \dfrac{2}{3 m_{\mu}^2 Q_1^2 \overline{Q}_1^6} \Bigg( -12 m_{\mu}^4 - Q_1^4 ( -1 + \sigma_1 ) + 2 m_{\mu}^2 Q_1^2 ( -9 + 10 \sigma_1 ) \\
&\quad + \overline{Q}_1 \left[ 6 m_{\mu}^2 Q_1 (3 - 2 \sigma_1 ) - 3 Q_1^3 ( -1 + \sigma_1 ) \right] y_{23} \\
&\quad + \left[ 48 m_{\mu}^4 + 10 Q_1^4 ( -1 + \sigma_1 ) + 4 m_{\mu}^2 Q_1^2 ( -9 + 4 \sigma_1 ) \right] y_{23}^2 \Bigg) \, ,
\end{aligned}
\end{equation}

% T10 \overline{\Pi}_{10}
\begin{equation}
\dfrac{T_{10} \overline{\Pi}_{10}}{\hat{\Pi}_{39}^{q_1,\mathrm{R}}} =  \dfrac{8}{3 m_{\mu}^2 \overline{Q}_1^6} \left( Q_1^2 ( -1 + \sigma_1 ) + m_{\mu}^2 ( -6 + 4 \sigma_1 ) \right) \left( 1 - y_{23}^2 \right) \, ,
\end{equation}

% T11 \overline{\Pi}_{11}
\begin{equation}
\begin{aligned}
\dfrac{T_{11} \overline{\Pi}_{11}}{\hat{\Pi}_{54}^{q_2,\mathrm{R}}} &= -\dfrac{8}{3 m_{\mu}^2 \overline{Q}_1^6} \Bigg( -Q_1^2 ( -1 + \sigma_1 ) + m_{\mu}^2 ( -6 + 8 \sigma_1 ) \\
&\quad + \left[ -2 Q_1^2 ( -1 + \sigma_1 ) + 4 m_{\mu}^2 \sigma_1 \right] y_{23}^2 \\
&\quad + \left[ -12 m_{\mu}^2 + 6 Q_1^2 ( -1 + \sigma_1 ) \right] y_{23}^4 \Bigg) \, ,
\end{aligned}
\end{equation}

% T12 \overline{\Pi}_{12}
\begin{equation}
\begin{aligned}
\dfrac{T_{12} \overline{\Pi}_{12}}{\hat{\Pi}_{54}^{q_1,\mathrm{R}}} &= \dfrac{4}{3 m_{\mu}^2 \overline{Q}_1^6} \Bigg( 5 Q_1^2 ( -1 + \sigma_1 ) + 2 m_{\mu}^2 ( -9 + 4 \sigma_1 ) \\
&\quad + \left[ 4 m_{\mu}^2 ( -3 + \sigma_1 ) + 4 Q_1^2 ( -1 + \sigma_1 ) \right] y_{23}^2 \\
&\quad + \left[ 12 m_{\mu}^2 - 6 Q_1^2 ( -1 + \sigma_1 ) \right] y_{23}^4 \Bigg) \, .
\end{aligned}
\end{equation}

For $q_2$ small

% T1 \overline{\Pi}_1
\begin{equation}
\dfrac{T_1 \overline{\Pi}_1}{\hat{\Pi}_1^{q_2,\mathrm{R}}} = 0 \, ,
\end{equation}

% T2 \overline{\Pi}_2
\begin{equation}
\dfrac{T_2 \overline{\Pi}_2}{\hat{\Pi}_1^{q_3,\mathrm{R}}} = -\dfrac{64}{m_{\mu}^2 \overline{Q}_2^6} \left( -Q_2^2 ( -1 + \sigma_2 ) + 2 m_{\mu}^2 \sigma_2 \right) \left( 1 - y_{31}^2 \right)
\, ,
\end{equation}

% T3 \overline{\Pi}_3
\begin{equation}
\begin{aligned}
\dfrac{T_3 \overline{\Pi}_3}{\hat{\Pi}_4^{q_2,\mathrm{R}}} &= -\dfrac{8}{3 m_{\mu}^2 Q_2^2 \overline{Q}_2^6} \Bigg( -12 m_{\mu}^4 - 5 Q_2^4 ( -1 + \sigma_2 ) + 2 m_{\mu}^2 Q_2^2 ( 3 + 2 \sigma_2 ) \\
&\quad + \overline{Q}_2 \left[ 3 Q_2^3 ( -1 + \sigma_2 ) + 6 m_{\mu}^2 Q_2 ( -3 + 2 \sigma_2 ) \right] y_{31} \\
&\quad + 8 \left[ 6 m_{\mu}^4 + m_{\mu}^2 Q_2^2 ( -3 + \sigma_2 ) + Q_2^4 ( -1 + \sigma_2 ) \right] y_{31}^2 \Bigg)
\, ,
\end{aligned}
\end{equation}

% T4 \overline{\Pi}_4
\begin{equation}
\dfrac{T_4 \overline{\Pi}_4}{\hat{\Pi}_4^{q_3,\mathrm{R}}} = 0
\, ,
\end{equation}

% T5 \overline{\Pi}_5
\begin{equation}
\begin{aligned}
\dfrac{T_5 \overline{\Pi}_5}{\hat{\Pi}_7^{q_2,\mathrm{R}}} &= \dfrac{4}{3 m_{\mu}^2 Q_2^2 \overline{Q}_2^6} \Bigg( -12 m_{\mu}^4 + 2 m_{\mu}^2 Q_2^2 (9 - 2 \sigma_2 ) - 7 Q_2^4 ( -1 + \sigma_2 ) \\
&\quad + \overline{Q}_2 \left[ 3 Q_2^3 ( -1 + \sigma_2 ) + 6 m_{\mu}^2 Q_2 ( -3 + 2 \sigma_2 ) \right] y_{31} \\
&\quad + \left[ 48 m_{\mu}^4 + 4 Q_2^4 ( -1 + \sigma_2 ) - 8 m_{\mu}^2 Q_2^2 \sigma_2 \right] y_{31}^2 \Bigg) \, ,
\end{aligned}
\end{equation}

% T6 \overline{\Pi}_6
\begin{equation}
\dfrac{T_6 \overline{\Pi}_6}{\hat{\Pi}_7^{q_3,\mathrm{R}}} = -\dfrac{12}{m_{\mu}^2 Q_2 \overline{Q}_2^6} \left( 2 m_{\mu}^2 - Q_2^2 ( -1 + \sigma_2 ) \right) y_{31} \left( 1 - y_{31}^2 \right) \, ,
\end{equation}

% T7 \overline{\Pi}_7
\begin{equation}
\dfrac{T_7 \overline{\Pi}_7}{\hat{\Pi}_7^{q_3,\mathrm{R}}} = -\dfrac{12}{m_{\mu}^2 Q_2 \overline{Q}_2^6} \left( 2 m_{\mu}^2 - Q_2^2 ( -1 + \sigma_2 ) \right) y_{31} \left( 1 - y_{31}^2 \right) \, ,
\end{equation}

% T8 \overline{\Pi}_8
\begin{equation}
\begin{aligned}
\dfrac{T_8 \overline{\Pi}_8}{\hat{\Pi}_{17}^{q_1,\mathrm{R}}} &= \dfrac{4}{3 m_{\mu}^2 Q_2^2 \overline{Q}_2^6} \Bigg( 12 m_{\mu}^4 + Q_2^4 ( -1 + \sigma_2 ) + 2 m_{\mu}^2 Q_2^2 ( -3 + 2 \sigma_2 ) \\
&\quad + 3 Q_2 \overline{Q}_2 \left[ m_{\mu}^2 (2 - 4 \sigma_2 ) + Q_2^2 ( -1 + \sigma_2 ) \right] y_{31} \\
&\quad + \left[ -48 m_{\mu}^4 + 4 Q_2^4 ( -1 + \sigma_2 ) + 8 m_{\mu}^2 Q_2^2 ( -3 + 2 \sigma_2 ) \right] y_{31}^2 \\
&\quad + \overline{Q}_2 \left[ 12 m_{\mu}^2 Q_2 - 6 Q_2^3 ( -1 + \sigma_2 ) \right] y_{31}^3 \\
&\quad + \left[ 4 Q_2^4 ( -1 + \sigma_2 ) + 8 m_{\mu}^2 Q_2^2 ( -3 + 2 \sigma_2 ) \right] y_{31}^4 \Bigg) \, ,
\end{aligned}
\end{equation}

% T9 \overline{\Pi}_9
\begin{equation}
\begin{aligned}
\dfrac{T_9 \overline{\Pi}_9}{\hat{\Pi}_{17}^{q_2,\mathrm{R}}} &= \dfrac{2}{3 m_{\mu}^2 Q_2^2 \overline{Q}_2^6} \Bigg( -12 m_{\mu}^4 - Q_2^4 ( -1 + \sigma_2 ) + 2 m_{\mu}^2 Q_2^2 ( -9 + 10 \sigma_2 ) \\
&\quad + \overline{Q}_2 \left[ 3 Q_2^3 ( -1 + \sigma_2 ) + 6 m_{\mu}^2 Q_2 ( -3 + 2 \sigma_2 ) \right] y_{31} \\
&\quad + \left[ 48 m_{\mu}^4 + 10 Q_2^4 ( -1 + \sigma_2 ) + 4 m_{\mu}^2 Q_2^2 ( -9 + 4 \sigma_2 ) \right] y_{31}^2 \Bigg) \, ,
\end{aligned}
\end{equation}

% T10 \overline{\Pi}_{10}
\begin{equation}
\dfrac{T_{10} \overline{\Pi}_{10}}{\hat{\Pi}_{39}^{q_2,\mathrm{R}}} = \dfrac{8}{3 m_{\mu}^2 \overline{Q}_2^6} \left( Q_2^2 ( -1 + \sigma_2 ) + m_{\mu}^2 ( -6 + 4 \sigma_2 ) \right) \left( 1 - y_{31}^2 \right) \, ,
\end{equation}

% T11 \overline{\Pi}_{11}
\begin{equation}
\begin{aligned}
\dfrac{T_{11} \overline{\Pi}_{11}}{\hat{\Pi}_{54}^{q_3,\mathrm{R}}} &= \dfrac{8}{3 m_{\mu}^2 Q_2^2 \overline{Q}_2^6} \Bigg( 12 m_{\mu}^4 + 2 m_{\mu}^2 Q_2^2 (3 - 4 \sigma_2 ) + Q_2^4 ( -1 + \sigma_2 ) \\
&\quad -12 m_{\mu}^2 Q_2 \overline{Q}_2 ( -1 + \sigma_2 ) y_{31} \\
&\quad + \left[ -48 m_{\mu}^4 + 6 m_{\mu}^2 Q_2^2 - 3 Q_2^4 ( -1 + \sigma_2 ) \right] y_{31}^2 \\
&\quad + \overline{Q}_2 \left[ 6 m_{\mu}^2 Q_2 - 3 Q_2^3 ( -1 + \sigma_2 ) \right] y_{31}^3 \\
&\quad + \left[ 2 Q_2^4 ( -1 + \sigma_2 ) + 4 m_{\mu}^2 Q_2^2 ( -3 + 2 \sigma_2 ) \right] y_{31}^4 \Bigg) \, ,
\end{aligned}
\end{equation}

% T12 \overline{\Pi}_{12}
\begin{equation}
\begin{aligned}
\dfrac{T_{12} \overline{\Pi}_{12}}{\hat{\Pi}_{54}^{q_2,\mathrm{R}}} &= \dfrac{4}{3 \overline{Q}_2^6} \Bigg( 18 - \dfrac{5 Q_2^2 ( -1 + \sigma_2 )}{m_{\mu}^2} - 8 \sigma_2 \\
&\quad + \left[ -4 ( -3 + \sigma_2 ) - \dfrac{4 Q_2^2 ( -1 + \sigma_2 )}{m_{\mu}^2} \right] y_{31}^2 \\
&\quad + \left[ -12 + \dfrac{6 Q_2^2 ( -1 + \sigma_2 )}{m_{\mu}^2} \right] y_{31}^4 \Bigg) \, .
\end{aligned}
\end{equation}
The integration is to be made over a symmetric interval in $y_{jk}$. Considering that the remaining integrand is symmetric in this variable, we observe how any potential $D=4+\delta$ contribution even in the $y_{jk}$ variable can only start at $T_i \overline{\Pi}_i\sim 1/\overline{Q}^{6+\delta}_i$, while any potential odd contribution could start at $T_i \overline{\Pi}_i\sim 1/\overline{Q}^{5+\delta}_i$.

\subsection{Combined expansion}

Let us now analyze the different studied contributions up to gluonic corrections. Purely from dimension $D=3$ we only have a term giving a contribution\footnote{Extension to NLO OPE is straightforward.}
\begin{equation}
\overline{\Pi}_2^{q_2} \to \hat{\Pi}_1^{q_3,\mathrm{R}}=\frac{2 \omega_L}{\pi^2}   \, .
\end{equation}
From the quark $D=4$ operators we have, for the $\overline{\Pi}_i$ that can contribute,
\begin{equation}
\overline{\Pi}_i^{q_1} \to \frac{\omega_T}{\pi^2} \left\{
\begin{aligned}
0 & \quad \text{for } i = 5, 7, 9 \\
1  & \quad \text{for } i = 3, 4 \\
4 & \quad \text{for } i = 6, 11 \\
-4 & \quad \text{for } i = 10, 12 \\
\end{aligned}
\right.\quad
\overline{\Pi}_i^{q_2} \to \frac{\omega_T}{\pi^2} \left\{
\begin{aligned}
0 & \quad \text{for } i =  2, 6, 7, 8, 9, 11 \\
1  & \quad \text{for } i = 3 \\
4 & \quad \text{for } i = 5, 12 \\
-4 & \quad \text{for } i = 10 \\
\end{aligned}
\right.
\end{equation}
and
\begin{equation}\label{eq:pibarsomplus}
\overline{\Pi}_i^{q_1} \to \pm 8 \frac{\omega^{(\pm)}}{Q_1^2} \left\{
\begin{aligned}
0 & \quad \text{for } i =  5, 7, 9  \\
1  & \quad \text{for } i = 3, 4  \\
4 & \quad \text{for } i = 6, 12 \\
-4 & \quad \text{for } i =  10, 11  \\
\end{aligned}
\right.\quad
\overline{\Pi}_i^{q_2} \to \pm 8 \frac{\omega^{(\pm)}}{Q_2^2} \left\{
\begin{aligned}
0 & \quad \text{for } i =  2, 6, 7, 8, 9, 11 \\
1  & \quad \text{for } i = 3 \\
4 & \quad \text{for } i = 5 \\
-4 & \quad \text{for } i = 10, 12
\end{aligned}
\right.
\end{equation}
Considering that they do not depend on $y_{jk}$ at the studied order, they can only contribute at order $1/\overline{Q}_i^6$. $\omega^{(+)}\equiv \omega_{+}^{D,1} + \omega_{+}^{D,2}$ and $\omega^{(-)}\equiv \omega_{+}^{D,1} - \omega_{+}^{D,4}$ give contributions to the separate $a_\mu^{\mathrm{HLbL},i}$ when using the splitting into corners made in Refs.~\cite{Colangelo:2015ama,Colangelo:2017fiz}. These contributions however exactly cancel in the sum. In other words, if we restore the original crossing of the integrand by taking 
$a_\mu^{\mathrm{HLbL}}=\frac{\sum_i^3 a_\mu^{\mathrm{HLbL},i}}{3}$  and relabeling dummy integration variables, we obtain that the integrand vanishes for these contributions. The same cancellation is observed for the gluonic corrections studied in section~\ref{sec:gluonic}. 

Remarkably, from the very many gluonic form factors (and also the remaining photon operator contribution), simply taking into account that at the studied order $T_1^{q_1}=T_2^{q_1}=T_{8}^{q_1}=T_{1}^{q_2}=T_{4}^{q_2}=0$ and $T_{6}^{q_2}=T_{7}^{q_2}$, only one contribution can potentially survive, $\omega^G_{\mathrm{eff}}\equiv \frac{(-\omega^G_6 -3\, \omega^G_7)}{12 \pi^2} $. However its contribution to the remaining $\overline{\Pi}_n$ is functionally identical to the one of $\omega^{(+)}$ given in Eq.~(\ref{eq:pibarsomplus}), and thus vanishes when summing over corners.

Summing the corner integrands ($F(Q_1^2,Q_2^2,Q_3^2)+F(Q_2^2,Q_3^2,Q_1^2)+F(Q_3^2,Q_1^2,Q_2^2)$ where $F$ represents the integrand) and relabeling dummy variables one miraculously finds, when summing over the $36=3\times 12$ functions, keeping only the even pieces in the $y$ variable (and thus also restoring full crossing of the integrand), 
\begin{align}\nonumber \label{eq:result}
\sum_i^{36} T_i\overline{\Pi}_i&=\frac{128}{3 m_{\mu}^2 \pi^2 \overline{Q}^6} \Big( -Q^2 (-1 + \sigma) \left[ 3 \omega_L (-1 + y^2) + \omega_T (2 + y^2) \right] \\&+ m_{\mu}^2 \left[ 6 \omega_L \sigma (-1 + y^2) - 2 \omega_T (-3 + 2 \sigma) (2 + y^2) \right] \Big) \, .
\end{align}
Let us note we do not have a formal proof that higher order terms in the tensor expansion cannot change this result. For this we would need a better understanding of the actual origin of these strong cancellations.
However, there are indications that suggest that this equation may give the leading term of the integrand, at least in the chiral limit. First, the $D=3$ operator does not contribute at order $1/\overline{Q}^4$ and $1/\overline{Q}^5$. Analogously, all the potential $D=4$ contributions studied here not related to the form factors already present at $D=3$, including gluonic form factors and gluonic corrections to quark operators, do not give any contribution at all either at $1/\overline{Q}^5$ or $1/\overline{Q}^6$, which anyway would be suppressed by extra powers of $\alpha_s(\overline{Q}_3)$. It seems reasonable to assume that any higher-order topology will follow the trend. Potential $D=5$ $\overline{\Pi}_n$ contributions even in $y_{jk}$ can only contribute at order $1/\overline{Q}^7_i$.

The remaining potential contributions from $D=5$ odd in $y_{jk}$, could possibly enter at $1/\overline{Q}^6_i$.\footnote{Let us note that for the symmetric $y_{jk}=0$ point they cannot contribute in any case.} Unfortunately, we do not have the form factor decomposition of the OPE all the way up to $D=5$. However we can extract their corresponding contributions from the two-loop results of Ref.~\cite{Bijnens:2022itw} in the perturbative regime, together with any remnant contribution coming from $D=3$ or $D=4$.  Not surprisingly we find that all potential contributions from the quark loop are zero and that all potential odd contributions in $y_{jk}$ from the gluonic corrections conspire to exactly cancel when summing over the different $\overline{\Pi}_n$, once again suggesting that Eq.~(\ref{eq:result}) is valid beyond the perturbative regime without any extra leading contribution. In that case, the form factors of the axial current (intuitively enough, the operator entering at leading order in the OPE) fully determine the leading term of the integrand in this kinematic regime.

\subsection{Validity domain of OPE result}
A natural question is how much we can push the inverse $\overline{Q}_i$ expansion. This depends on many higher dimensional nonperturbative form factors, which we do not know precisely in general. However we do know them when the matrix elements can be computed perturbatively and one may expect the validity domain of the expansions to be similar. 

\begin{figure}[tbh]
\includegraphics[height=0.32\textwidth]{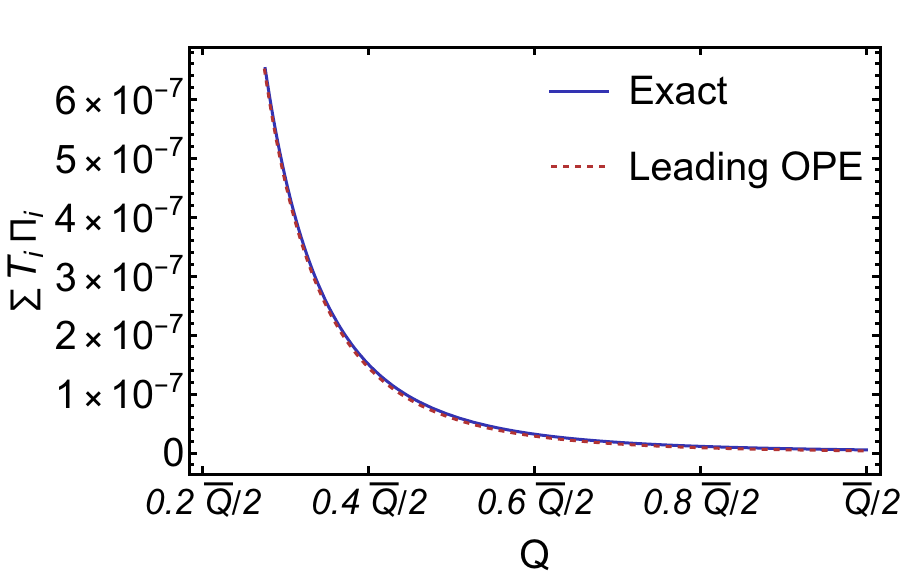}
\includegraphics[height=0.32\textwidth]{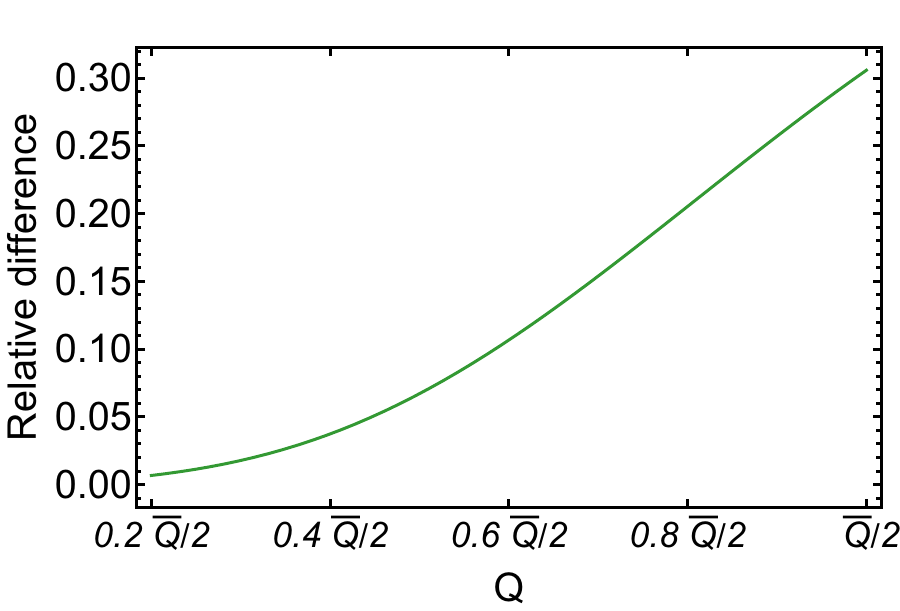}\\
\includegraphics[height=0.32\textwidth]{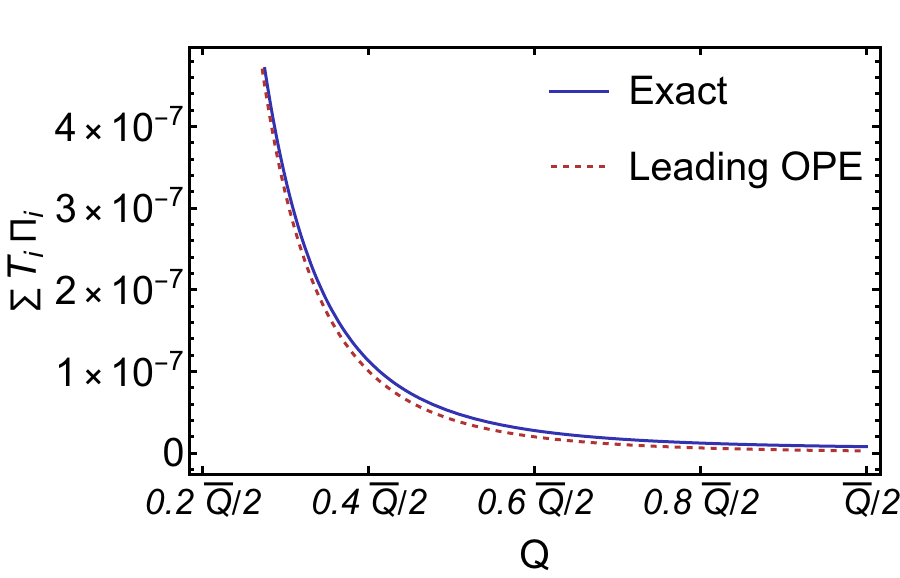}
\includegraphics[height=0.32\textwidth]{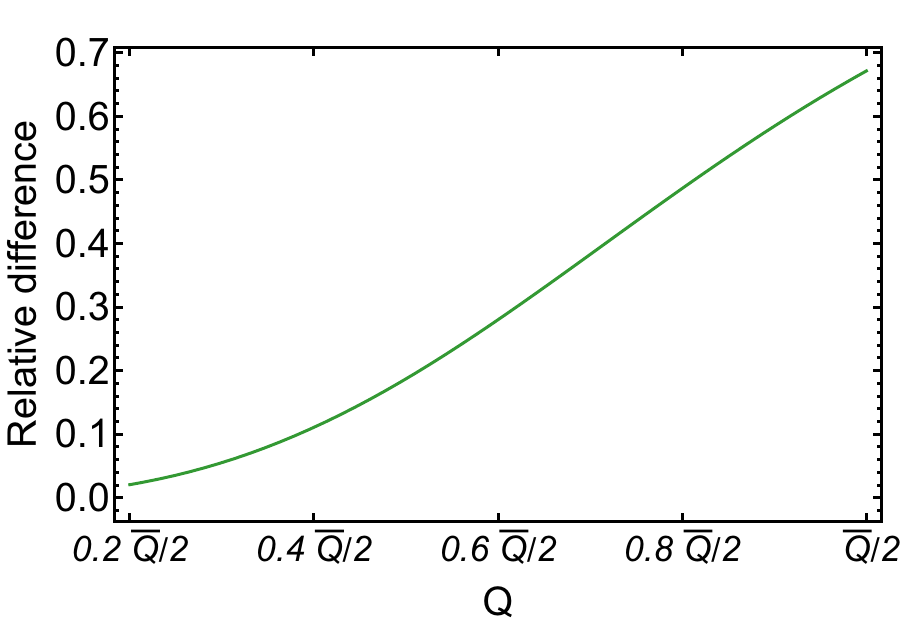}
\label{fig:result}
\caption{Expansion of $\sum T_i \Pi_i$ in $\sum_q N_c e_q^4$ units adding the three corners and symmetrizing in the $y$ variable (i.e. restoring crossing) using the exact expression with quark loop ansatz and Eq.~(\ref{eq:result}) fixing $\overline{Q}=10\,\mathrm{GeV}$. Top panel takes $y=0$, bottom panel $y=0.7$.}
\end{figure}

Let us then take on the one hand the (kilometric) ``exact" ansatz for the $\sum_i T_i \overline{\Pi}_i$, taking the quark loop for the $\overline{\Pi}_i$. Let us take, on the other hand, Eq.~(\ref{eq:result}) with the corresponding $2\omega_T=\omega_L=-2/Q_i^2$ (in $N_c\sum e_q^4$ units). As benchmark points we fix $\overline{Q}_{i}=10 \, \mathrm{GeV}$. The result of $Q_i$ from $\overline{Q}/10$ to the symmetric point $\overline{Q}/2$ for two values of $y$ is displayed in Figure~\ref{fig:result}. We thus find that the leading term, corresponding to the axial current form factor contributions, offers a reasonable description up to near the point in which $Q$ is as large as, at least, one of the other $Qs$.

\section{Conclusions and outlook}\label{sec:conclusions}
In this paper we derived short-distance constraints on the hadronic light-by-light contribution to the muon $g-2$. We considered constraints for the corner (Melnikov-Vainshtein) kinematics of the underlying hadronic light-by-light tensor, extending previous works~\cite{Melnikov:2003xd,Bijnens:2022itw} using operator product expansion techniques developed in Refs.~\cite{Bijnens:2019ghy,Bijnens:2020xnl,Bijnens:2021jqo,Bijnens:2022itw}. We added the gluonic corrections to the OPE\footnote{The exception is the $\alpha_s$ correction to the gluonic operator in the nonperturbative regime since there diagrams involving a triple gluon vertex can contribute.} and to the perturbative matrix elements.
In the kinematical limit where all three momenta are large, we showed how our short-distance constraints from Ref.~\cite{Bijnens:2022itw} and those including perturbative order-$\mathcal{\alpha}_s$ corrections derived here reproduce our old results from Ref.~\cite{Bijnens:2021jqo}. Our calculations were performed in the chiral limit, which in future work will be extended to include chiral corrections. 

We discussed how the matrix elements of the operators can be written in general form factors~\cite{Bijnens:2022itw}, and by using the relations derived from the absence of kinematical singularities and expanding the coefficient functions $T_i$ in the corners, how the entire result for the final $g-2$ integral to the order considered can be rewritten in the longitudinal and transverse form factors associated to the axial current, as exemplified in Eq.~(\ref{eq:result}).
In the future it will be important to combine the short-distance constraints derived here and in Refs.~\cite{Melnikov:2003xd,Bijnens:2019ghy,Bijnens:2020xnl,Bijnens:2021jqo,Bijnens:2022itw} with data-driven evaluations of the hadronic light-by-light contribution beyond what was done in the White Paper~\cite{Aoyama:2020ynm}, in order to reach the target relative precision of $10\%$ needed to match the precision goal of the Fermilab experiment~\cite{Muong-2:2015xgu,Muong-2:2023cdq,Muong-2:2021ojo}.

\section*{Acknowledgements}
J.~B.~is supported by the Swedish
Research Council grants contract numbers 2016-05996 and 2019-03779. 
N.~H.-T.~is funded by the UK Research and Innovation, Engineering and Physical Sciences Research Council, grant number EP/X021971/1. N.~H.-T.~wishes to thank Lund University for hosting him while parts of the project were completed. 
A.~R.-S.~is funded in part by MIUR contract number 2017L5W2PT and by the Generalitat Valenciana (Spain) through the plan GenT program (CIDEIG/2023/12) and 
by the Spanish Government (Agencia Estatal de Investigación MCIN/AEI/10.13039/501100011033) Grants No. PID2020–114473GB-I00 and No. PID2023-146220NB-I00

\appendix

\section{Kinematics and integration over different variables}\label{app:kinematics}

In this appendix we discuss a number of features of the various choices of variables and integration regions. The simplest variables are $Q_1,Q_2,Q_3$. In terms of these the integration region is given by
\begin{align}
\int_0^\infty dQ_1
\int_0^\infty dQ_2
\int_{\vert Q_1-Q_2\vert}^{Q_1+Q_3} dQ_3 \, .
\end{align}
The boundaries on $Q_3$ follow from the requirement that the three can be the sides of a triangle.
The integral is fully symmetric under permutations of $Q_1,Q_2,Q_3$. 

\begin{figure}
\centerline{\includegraphics[width=0.49\textwidth]{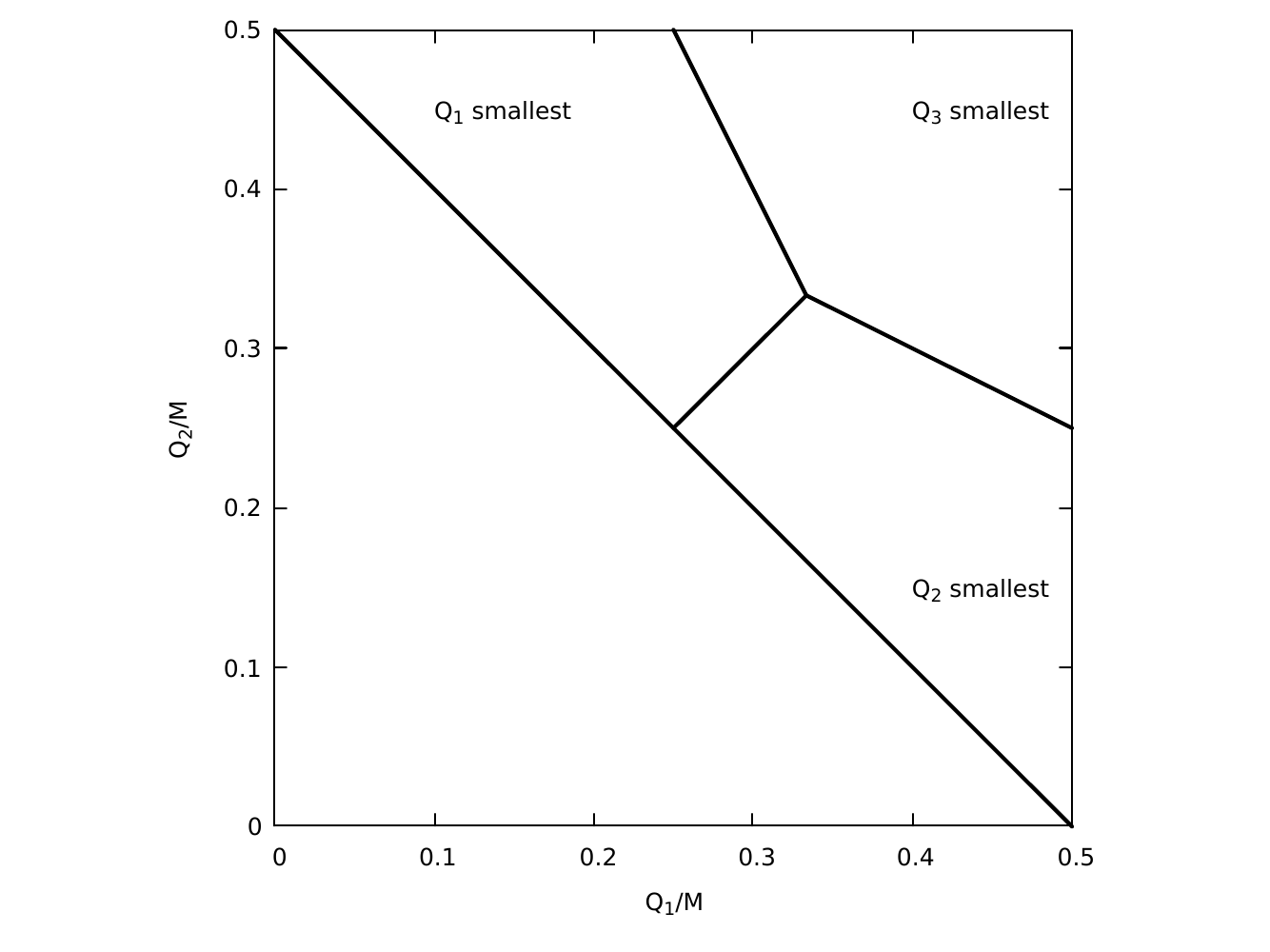}}
    \caption{The allowed region for $Q_1+Q_2+Q_3=M$ at constant $M$ is the upper triangle. Also indicated are the regions corresponding to the smallest $Q_i$.}
    \label{fig:regionssymmetric}    
\end{figure}

To visualize the regions discussed below we fix the sum of all three $Q_i$, $M=Q_1+Q_2+Q_3$. The allowed region is shown in the $Q_1,Q_2$ plane in Fig.~\ref{fig:regionssymmetric}.

Alternatively we define the corner variables $\overline Q_3=Q_1+Q_2$, $Q_3$ and $y_{12}=\dfrac{Q_1-Q_2}{Q_3}$. In these variables the integration becomes
\begin{align}
\frac{1}{2}\int_0^\infty d\overline Q_3 \int_0^{\overline Q_3}dQ_3\,Q_3\int_{-1}^1 dy_{12} \, .
\end{align}
The cyclic permutations $\overline Q_2=Q_3+Q_1, Q_2, y_{31}=\dfrac{Q_3-Q_1}{Q_2}$
and $\overline Q_1=Q_2+Q_3, Q_1, y_{23}=\dfrac{Q_2-Q_3}{Q_1}$ are other possibilities.

In this paper we discuss expansions in $Q_i/\overline Q_i$ and this can be done by using the three regions shown in Fig.~\ref{fig:regionssymmetric}. 
This corresponds to using
\begin{align}
    \label{eq:smallQi}
\frac{1}{2}\int_0^\infty d\overline Q_3 \int_0^{\overline Q_3/2}dQ_3\,Q_3\int_{-y_{12\textrm{max}}}^{y_{12\textrm{max}}}+\textrm{cyclic} \, ,
\end{align}
with $y_{12\textrm{max}}=\min\left(1,\dfrac{\overline Q_3-2 Q_3}{Q_3}\right)$, i.e. 1 for $Q_3 \le \dfrac{\overline Q_3}{3}$ and $y_{12\textrm{max}}=\dfrac{\overline Q_3-2 Q_3}{Q_3}$ for $Q_3 \ge \dfrac{\overline Q_3}{3}$. The three terms in Eq.~(\ref{eq:smallQi}) correspond to the three regions in Fig.~\ref{fig:regionssymmetric}.

\subsection{Cut-off in $Q_1,Q_2,Q_3$}

\begin{figure}
    \begin{minipage}{0.49\textwidth}
        \includegraphics[width=0.99\textwidth]{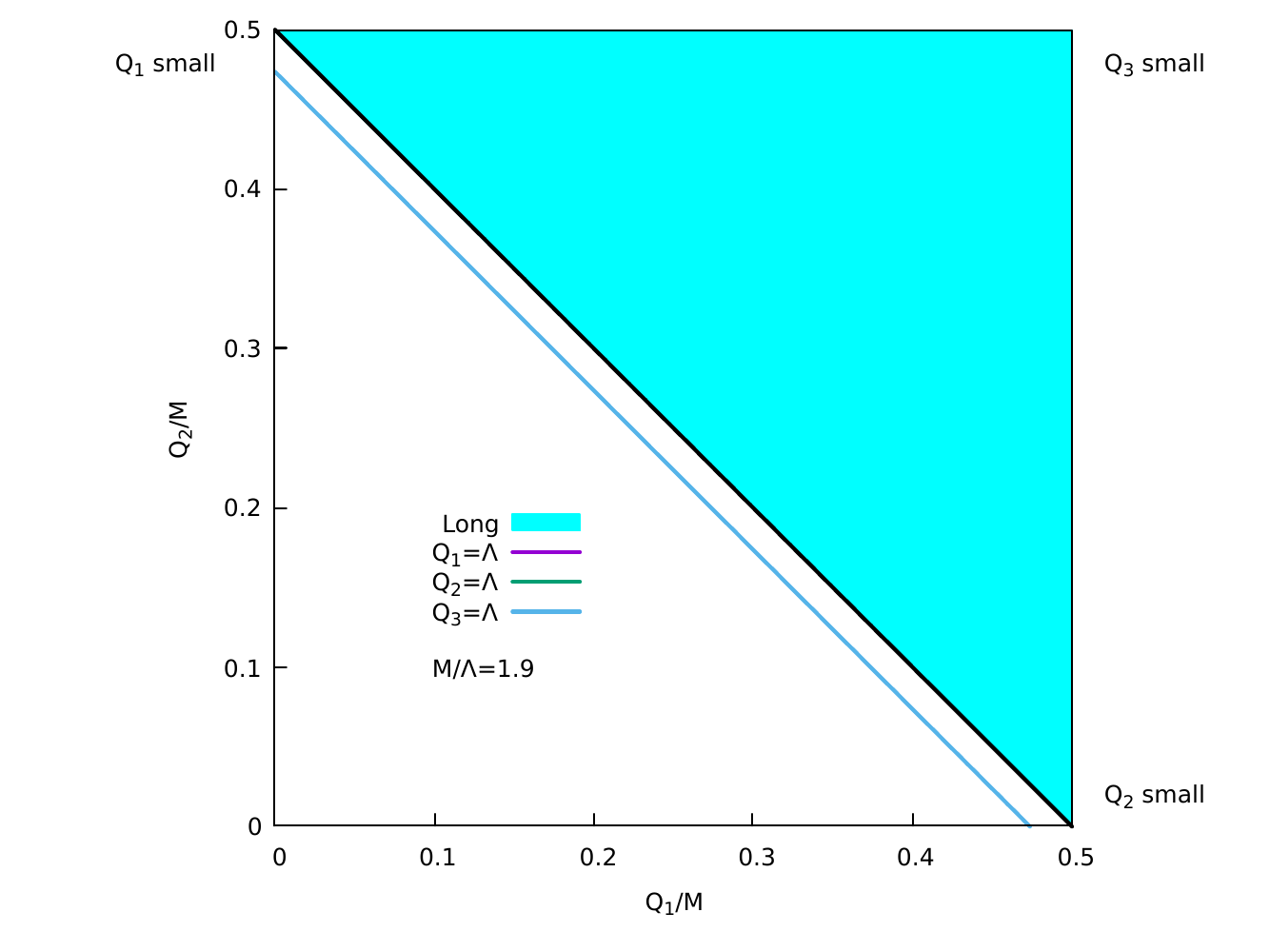}
        \centerline{(a)}
    \end{minipage}  
        \begin{minipage}{0.49\textwidth}
            \includegraphics[width=0.99\textwidth]{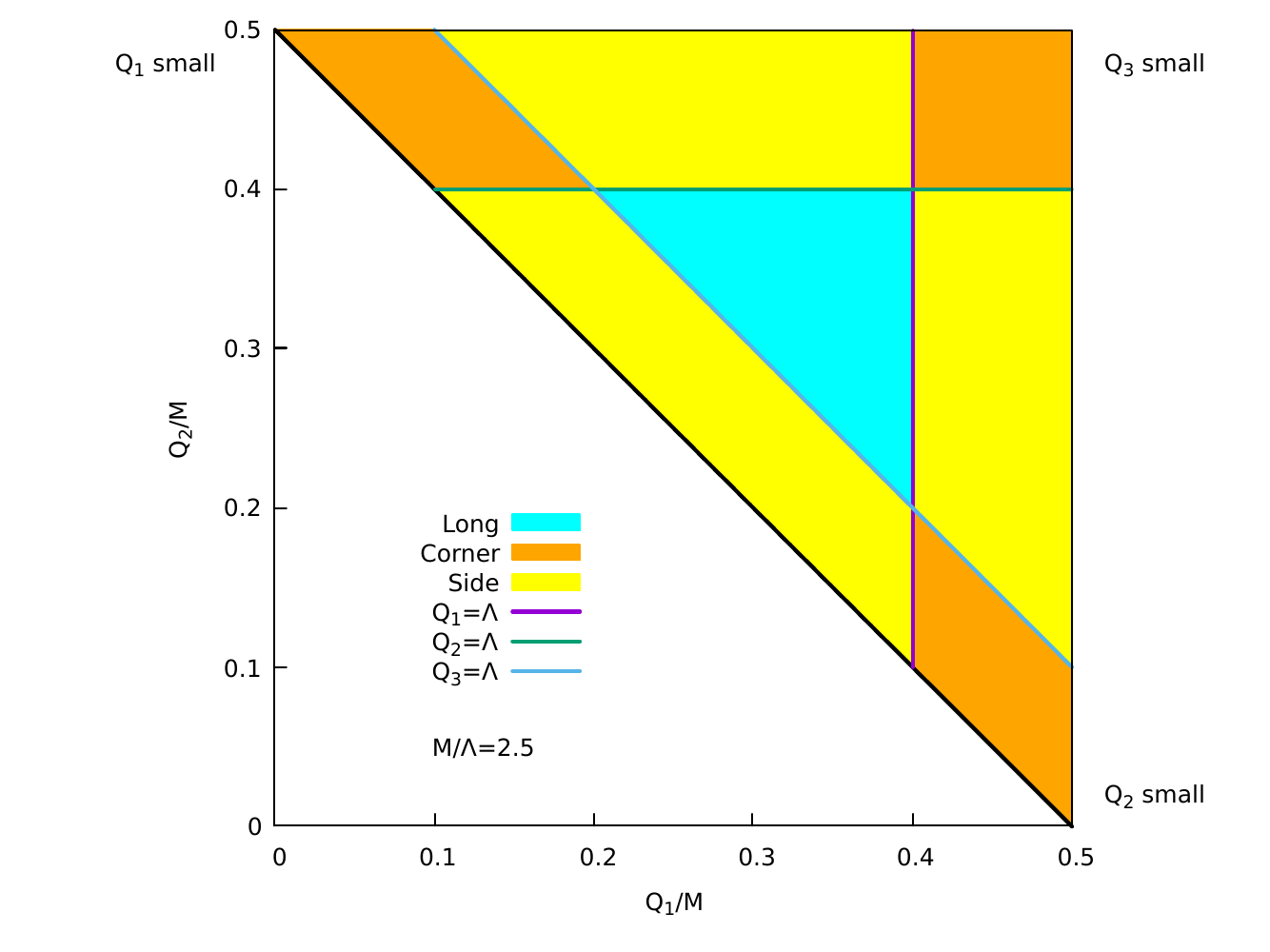}
            \centerline{(b)}
    \end{minipage}  
    \begin{minipage}{0.49\textwidth}
        \includegraphics[width=0.99\textwidth]{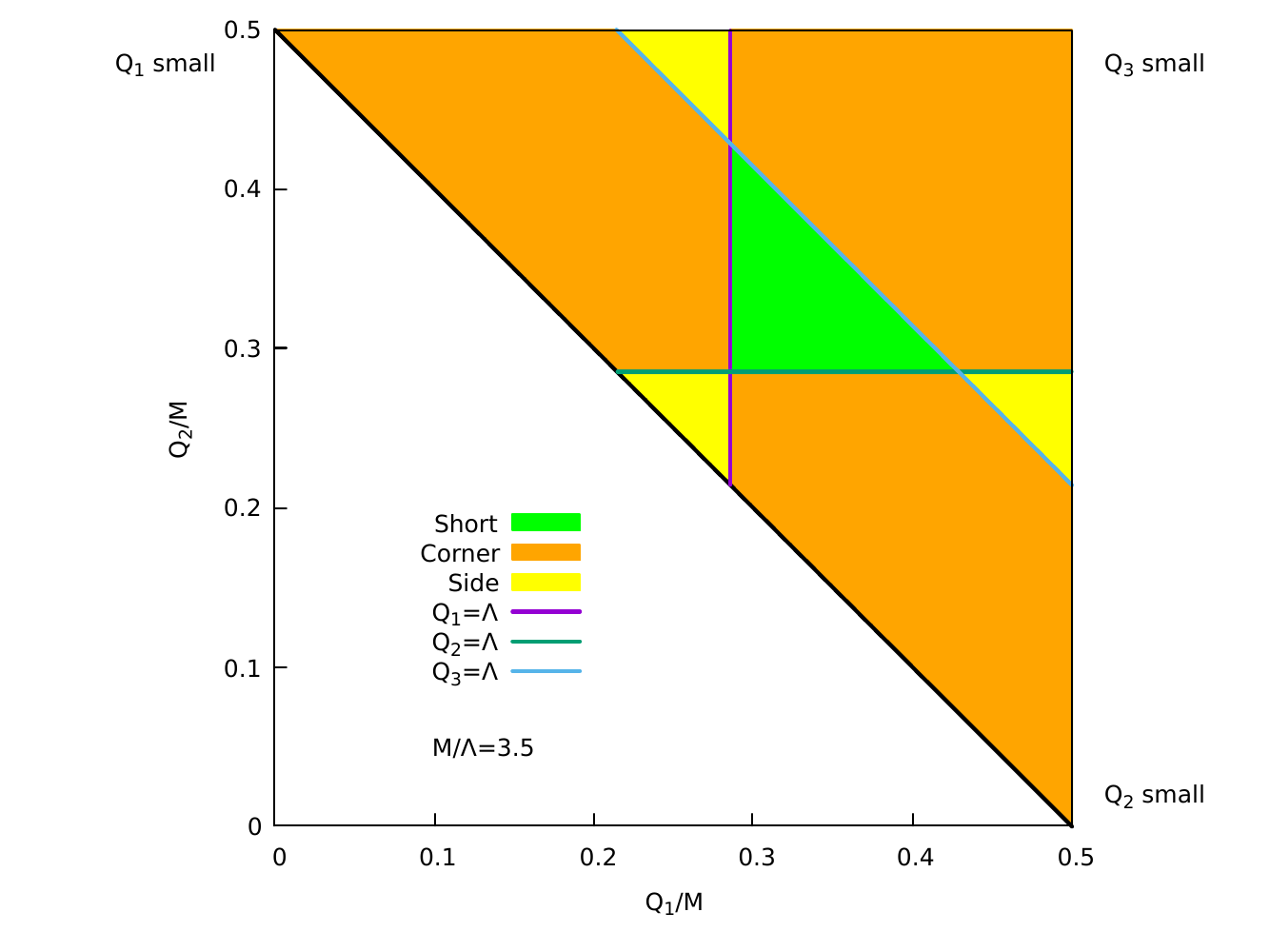}
        \centerline{(c)}
    \end{minipage}  
    \begin{minipage}{0.49\textwidth}
        \includegraphics[width=0.99\textwidth]{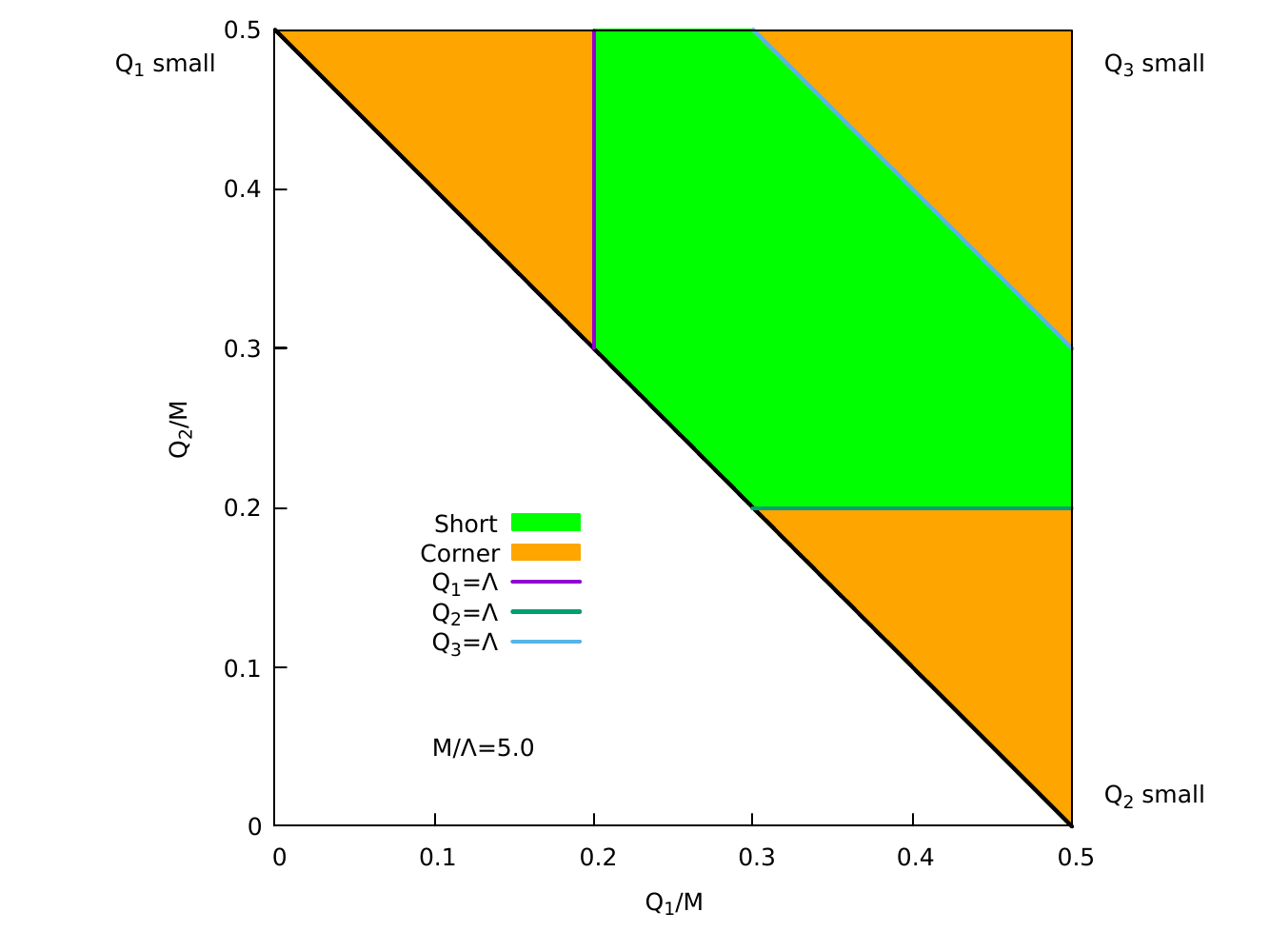}
        \centerline{(d)}
    \end{minipage}  
    \caption{The different regions Long, Short, Side and Corner for the $Q_1+Q_2+Q_3=M$ for $M/\Lambda=1.9,2.5,3.5,5$}  
    \label{fig:regions}
\end{figure}

If we have a cut-off (or dividing scale) $\Lambda$ we have eight different regions depending on whether the $Q_i$ are larger or smaller than $\Lambda$:
\begin{itemize}
\item Purely long distance (Long): 
\begin{align}
Q_1,Q_2,Q_3 < \Lambda ,
\end{align}
\item Mixed with two smaller (Side):
\begin{align}
    Q_1,Q_2 <& \Lambda < Q_3,&
    Q_2,Q_3 <& \Lambda < Q_1,&
    Q_3,Q_1 <& \Lambda < Q_2,&
\end{align}
\item Mixed with two larger (Corner)
\begin{align}
    Q_1 <& \Lambda < Q_2,Q_3,&
    Q_2 <& \Lambda < Q_3,Q_1,&
    Q_3 <& \Lambda < Q_1,Q_2,&
\end{align}
\item Purely short-distance (Short)
\begin{align}
    Q_1,Q_2,Q_3 > \Lambda ,
\end{align}
\end{itemize}  
We show how they look like for $Q_1+Q_2+Q_3=M$. For $M\le2\Lambda$ only Long is possible;
for $2\Lambda\le M\le3\Lambda$ Long, Side and Corner occur;
for $3\Lambda\le M\le4\Lambda$ Side, Corner and Short occur and for $4\Lambda \le M$ we have Corner and Short. The four cases are shown in Fig.~\ref{fig:regions}

\subsection{Cut-off in $\overline Q_3,Q_3,y_{12}$}

The various regions separated by a cut-off $\Lambda$ are a bit complicated in the other variables.
We define $\tilde q_3=\vert \overline Q_3-2\Lambda\vert$ and $y_\textrm{max} = \tilde q_2/Q_3$.
The resulting corresponding regions are given in Table~\ref{tab:regions}.

\begin{table}[tb!]
\renewcommand{\arraystretch}{1.2}
\newcommand{\qt}{\ensuremath{\tilde q_3}}    
\newcommand{\QT}{\ensuremath{\overline Q_3}}
\newcommand{\ym}{\ensuremath{y_\textrm{max}}}
\begin{center}
\begin{tabular}{|c|c c|c c|c c|}
\hline 
& \multicolumn{2}{c|}{$d$\QT}    
& \multicolumn{2}{c|}{$dQ_3$}    
& \multicolumn{2}{c|}{$dy_{12}$}\\    
\hline
\multirow{3}{7em}{$Q_1,Q_2,Q_3 \le \Lambda$} & 0 & $\Lambda$ & 0 & \QT & $-1$ & 1\\
\cline{2-7}
    & \multirow{2}{1.4em}{$\Lambda$} & \multirow{2}{1.4em}{$2\Lambda$}
        & 0 & \qt & $-1$ & $1$\\
    & &    & \qt & \QT & $-\ym$ & \ym\\
\hline
$Q_2,Q_3\le \Lambda \le Q_1$ & $\Lambda$ & $3\Lambda$ & \qt & $\Lambda$ & \ym & 1\\\hline
$Q_1,Q_3\le \Lambda \le Q_2$ & $\Lambda$ & $3\Lambda$ & \qt & $\Lambda$ & $-1$ & $-\ym$\\\hline
\multirow{3}{8em}{$Q_3\le\Lambda\le Q_1,Q_2$} & \multirow{2}{1.4em}{$2\Lambda$} &  \multirow{2}{1.4em}{$3\Lambda$} & 0 & \qt & $-1$ & $1$\\
 & & & \qt & $\Lambda$ & $-\ym$ & \ym\\
 \cline{2-7}
 & $3\Lambda$ & $\infty$ & 0 & $\Lambda$ & $-1$ & 1\\\hline
 $Q_1,Q_2\le \Lambda \le Q_3$ & $\Lambda$ & $2\Lambda$ & $\Lambda$ & \QT & $-\ym$ & \ym \\\hline
 \multirow{2}{8em}{$Q_2\le\Lambda\le Q_1,Q_3$} & $\Lambda$ & $3\Lambda$ & $\Lambda$ & \QT & \ym & 1\\
   & $3\Lambda$ & $\infty$ & \qt & \QT & \ym & 1\\\hline
   \multirow{2}{8em}{$Q_1\le\Lambda\le Q_2,Q_3$} & $\Lambda$ & $3\Lambda$ & $\Lambda$ & \QT & $-1$ & $-\ym$\\
   & $3\Lambda$ & $\infty$ & \qt & \QT & $-1$ & $-\ym$\\\hline
\multirow{3}{7em}{$\Lambda\le Q_1,Q_2,Q_3$} & $2\Lambda$ & $3\Lambda$ & $\Lambda$ & \QT & $-\ym$ & \ym\\
\cline{2-7}
 & \multirow{2}{1.4em}{$3\Lambda$}&\multirow{2}{1.4em}{$\infty$} & $\Lambda$ & \qt & $-1$ & 1\\
  & & & \qt & \QT & $-\ym$ & \ym\\\hline   
\end{tabular}
\end{center}
\caption{The different integration regions in $Q_1,Q_2,Q_3$ translated into regions for $\QT$, $Q_3$ and $y_{12}$. $\qt=\vert 2\Lambda-\QT\vert$ and $\ym=\qt/Q_3$.}
\label{tab:regions}
\renewcommand{\arraystretch}{1.}
\end{table}

\section{Connection with the axial current form factors}\label{app:axialcurrent}
As argued above, simply using the EOM and $\gamma$ algebra one finds at $d=4$
\begin{equation}
\mathcal{O}^{\alpha\beta}_{(4),A}=\frac{1}{4}\left(\gamma^\alpha\gamma^\beta\gamma^\gamma-\gamma^\gamma\gamma^\beta\gamma^\alpha\right)\left(\overrightarrow{D}_\gamma+\overleftarrow{D}_\gamma\right) \, .
\end{equation}
Using now that
\begin{equation}
\gamma^{\alpha}\gamma^{\beta}\gamma^{\gamma}-\gamma^{\gamma}\gamma^{\beta}\gamma^{\alpha}=-2i\, \epsilon^{\delta\alpha\beta\gamma}\gamma_\delta \gamma_5 \, ,
\end{equation}
and that
\begin{equation}
\lim_{q_4\to 0}i \partial_{q_4}^{\nu_4} \langle 0 | \partial_ {\gamma} (\bar{q}\gamma_{\delta}\gamma_5 q)   | \gamma^{\mu_3}\gamma^{\mu_4}\rangle=-iq_{3\,\gamma} \lim_{q_4\to 0}i \partial_{q_4}^{\nu_4} \langle 0 |  \bar{q}\gamma_{\delta}\gamma_5 q  | \gamma^{\mu_3}\gamma^{\mu_4}\rangle \, ,
\end{equation}
one finds the relation
\begin{equation}
\lim_{q_4\to 0}i \partial_{q_4}^{\nu_4} \langle 0 | \mathcal{O}^{\alpha\beta}_{(4),A}   | \gamma^{\mu_3}\gamma^{\mu_4}\rangle=-\frac{i}{2}\epsilon^{\delta\alpha\beta\gamma}(-iq_{3\, \gamma}) (i\partial_{q_4}^{\nu_4})\langle 0 | \bar{q}\gamma_{\delta}\gamma_5 q |\gamma^{\mu_3}\gamma^{\mu_4}\rangle \, ,
\end{equation}
valid for any flavor. In our case we want to convolute this with the following flavor structures
\begin{equation}
\frac{1}{e^2\hat{q}^2} \, i\sum_j \underbrace{\left(e_{q_j}^2  -\sum_k\frac{e_{q,k}^2}{3}\right)}_{e_{q_j}/3}  \, 
\lim_{q_4 \rightarrow 0}\, \partial ^{\nu_4}_{q_4} \, \Big\langle \mathcal{O}^{\alpha\beta}_{(4)} \Big\rangle ^{j, \mu_3,\, \mu_4}_{\overline{\textrm{MS}}(\mu)}=\sum_{i=1}^6 \omega_{(8)}^{D,i}\, L_i^{\alpha\beta\mu_3\mu_4\nu_4} \, , \label{eq:octetME2}
\end{equation}
\begin{equation}
\frac{1}{e^2\hat{q}^2}\, i\sum_j   \sum_k\frac{e_{q_k}^2}{3}    \lim_{q_4 \rightarrow 0}\partial^{\nu_4}_{q_4}\Big\langle \mathcal{O}^{\alpha\beta}_{(4)}\Big\rangle ^{j, \mu_3,\, \mu_4}_{\overline{\textrm{MS}}(\mu)}
=
\sum_{i=1} ^{6}\omega_{(1)}^{D,i} \, L_i^{\alpha\beta\mu_3\mu_4\nu_4} \, . 
\end{equation}
From Ref.~\cite{Bijnens:2022itw} we have
\begin{align}\nonumber
&\sum_{j}\frac{e_{q,j}^2}{e^2}\lim_{q_4 \rightarrow 0}\, i\, \partial_{q_4}^{\nu_4}\langle  0|\bar{q}_{j} \gamma_{\delta}\gamma_5 q_{j}|\gamma^{\mu_3}(q_3)\gamma^{\mu_4}(q_4)\rangle
\\
&=
-\frac{q_{3}^2}{4\pi^2}\, g_{\sigma \delta}\, \left(\epsilon^{\mu_3\mu_4\nu_4\sigma}\, \omega_{T}(q_3^{2})
-\frac{1}{q_3^2}\, \epsilon ^{q_3\mu_4\nu_4\sigma}\, q_{3}^{\mu_3}\, \omega_{T}(q_3^{2})
+\frac{1}{q_3^2}\, \epsilon ^{\mu_3 \mu_4 \nu_4 q_3}\, q_{3}^{ \sigma}\, \left[\omega_{L}(q_3^2)-\omega_{T}(q_3^2)\right] \right) \, .
\end{align}
with identical octet and singlet decomposition. Replacing one finds
\begin{equation}
\sum_{j}\frac{e_{q,j}^2}{e^2}\lim_{q_4\to 0}i \partial_{q_4}^{\nu_4} \langle 0 | \mathcal{O}^{\alpha\beta}_{(4),A}   | \gamma^{\mu_3}\gamma^{\mu_4}\rangle=\frac{\omega_T(q_3^2)}{8\pi^2}q_{3}^2 q_{3\,\gamma}\left(g^{\mu \mu_3}-\frac{q_{3}^{\mu} q_{3}^{\mu_3}}{q_3^2}\right)g^{\sigma \mu _4}\, \delta_{\sigma \nu_4\mu}^{\alpha\beta\gamma} \, ,
\end{equation}
where $\delta^{\alpha\beta\gamma}_{\sigma\nu_4\mu}=g^{\alpha}_{\sigma}g^{\beta}_{\nu_4}g^{\gamma}_{\mu}-\cdots=-\epsilon^{\delta\alpha\beta\gamma}\epsilon_{\delta\sigma \nu_4\mu}$.

%%%%%%%%%%%%%%%%%%%%%%Bibliography%%%%%%%%%%%%%%%%%%%%%%%%%%

\bibliographystyle{JHEP}
\bibliography{refs}

\end{document}